\documentclass[aps,prd,reprint,amsmath,amssymb,nofootinbib,longbibliography,floatfix]{revtex4-2}
\usepackage{graphicx,bm}
\usepackage{tikz}
\usepackage[colorlinks=true,linkcolor=blue,citecolor=blue,urlcolor=blue]{hyperref}
\hypersetup{pdftitle={The QCD phase diagram and critical end point at finite magnetic field from holography with neural networks},pdfauthor={Shaojie Liu, Yehong Zhang, Hui Liu, Xuanmin Cao},pdfsubject={PRD working revision: unified Bali background and recalibrated magnetic results}}
\usetikzlibrary{positioning,arrows.meta}
\begin{document}
\title{The QCD phase diagram and critical end point at finite magnetic field from holography with neural networks}
\author{Shaojie Liu}
\author{Yehong Zhang}
\author{Hui Liu}
\email{tliuhui@jnu.edu.cn}
\author{Xuanmin Cao}
\thanks{Corresponding author.}
\email{caoxm@jnu.edu.cn}
\affiliation{Department of Physics, Jinan University, Guangzhou 510632, China}
\date{October 6, 2026}
\begin{abstract}
We study the magnetic-field-driven critical end point (CEP) of the
QCD thermal transition at zero baryon chemical potential using a
five-dimensional Einstein--Maxwell--dilaton (EMD) model.
We calibrate the model to lattice entropy at zero and finite magnetic
fields using a neural-network formulation of holographic radial evolution.
As the magnetic field increases, the thermal transition changes
from a crossover to a first-order transition.
The merger of temperature turning points, a divergent specific heat,
and a vanishing squared sound speed identify
$(eB_c,T_c)\simeq(7.22\,\mathrm{GeV}^{2},95.9\,\mathrm{MeV})$.
The transition-region inflection of $s/T^3$ defines the crossover
temperature, while an equal-pressure Maxwell construction determines
the first-order coexistence line. The two lines meet at the CEP.
The change of transition order agrees qualitatively with lattice results.
The phase boundary is a prediction of the calibrated model.
\end{abstract}
\maketitle
\section{Introduction}
Strong magnetic fields can modify the thermodynamics and phase structure of
QCD matter~\cite{Andersen:2014xxa,Kharzeev:2012ph,Rougemont:2023gfz,
Miransky:2015ava}. They
are therefore part of the study of QCD under extreme conditions
~\cite{Fukushima:2010bq,Ratti:2018ksb,Braun-Munzinger:2015hba}. Primordial
magnetic fields may have formed during cosmological phase transitions
~\cite{Vachaspati:1991nm,Grasso:2000wj,Durrer:2013pga}. Strong transient
fields are also generated in noncentral heavy-ion collisions at RHIC and the
LHC~\cite{Skokov:2009qp,Voronyuk:2011jd,Deng:2012pc,Bzdak:2011yy,
McLerran:2013hla,Huang:2015oca}. Both settings motivate studies of QCD
thermodynamics at finite magnetic field. Fields of several
$\mathrm{GeV}^{2}$ probe a regime beyond conventional heavy-ion collision
conditions. In this regime, the deconfinement transition may change from a
crossover to a first-order transition. The two regimes would then meet at a
magnetic-field-driven critical end point (CEP).

At $\mu_B=0$ and $eB=0$, lattice QCD with physical quark masses finds a
smooth deconfinement crossover
~\cite{Aoki:2006we,Bhattacharya:2014ara,Borsanyi:2010bp,Aoki:2009sc,
HotQCD:2019xnw,Bazavov:2011nk}. The zero-field equation of state is also known with good
accuracy~\cite{HotQCD:2014kol,Borsanyi:2013bia,Bazavov:2017dus}.
The transition remains an analytic crossover at \(eB=3.25~\mathrm{GeV}^2\), whereas simulations of the asymptotically strong-field limit provide evidence for a first-order transition~\cite{Endrodi:2015oba}. A later calculation reports crossover
behaviour at $eB=4\,\mathrm{GeV}^{2}$ and a first-order transition at
$eB=9\,\mathrm{GeV}^{2}$~\cite{DElia:2021yvk}. It places the change of order,
and hence the CEP at $\mu_B=0$, within
$4<eB_c<9\,\mathrm{GeV}^{2}$. The location of the CEP inside this interval
is still unknown.


Gauge/gravity duality relates certain strongly coupled theories to classical
gravity~\cite{Maldacena:1997re,Gubser:1998bc,Witten:1998qj}. Bottom-up
nonconformal black-hole models reproduce major features of the QCD equation
of state~\cite{Gubser:2008ny,Gubser:2008yx,Gursoy:2007cb,Gursoy:2009jd}.
Dynamical holographic QCD and improved soft-wall models have also been
developed to describe chiral symmetry breaking and confinement
~\cite{Li:2012ay,LiHuang2013Spectra,Chelabi2016SoftWall}. EMD models have
also been used to obtain finite-$\mu_B$ phase diagrams with a CEP
~\cite{DeWolfe:2010he,Knaute:2017opk,Critelli:2017oub,Grefa:2021qvt,
Cai:2012xh,Cai:2022omk,LiEtAl2018CEP,ChenEtAl2019Criticality,
ZhaoEtAl2024TwoFlavor,JokelaEtAl2024Refining}.
Bayesian calibration of EMD models to lattice entropy and baryon-number
susceptibility at $\mu_B=0$ has also been used to infer the finite-density
critical-point location and its uncertainty~\cite{Hippert:2023bel}.
Dynamical and soft-wall
models give related results for the quarkyonic phase and finite-density
CEP~\cite{Chen:2019rez}, as well as pion condensation and pion
quasiparticles at finite isospin chemical potential
~\cite{Cao:2020isospin,Cao:2020pion}. The finite-$\mu_B$
CEPs discussed above are driven by baryon density rather than magnetic
field. The same holographic thermodynamic methods can be applied to a
magnetic CEP. At finite magnetic field, anisotropic EMD models reproduce
the field dependence of the crossover temperature~\cite{Rougemont:2015oea}.
Hot and magnetized holographic models also describe the chiral transition
and neutral-pion spectrum~\cite{Wen:2024pion}. Holographic calculations
have also studied inverse magnetic catalysis, the chiral transition, and
systems with both magnetic field and chemical potential
~\cite{Gursoy:2016ofp,Bohra:2020qom,Li:2016gfn,Fang:2016qss,Chen:2021mpt,
EvansEtAl2016IMC,GursoyEtAl2018BMu,BohraEtAl2020Magnetic}.

At \(\mu_B=0\), a magnetised EMD model predicts a first-order line ending at a magnetic CEP~\cite{Cao:2024jgt}. Related EMD studies have examined the combined dependence on magnetic field and chemical potential~\cite{Zhu:2023mfd}. At strong field, this result depends on the
magnetic backreaction encoded in $f(\phi)$. Earlier work fixed $V(\phi)$ from
the zero-field equation of state. The function $f(\phi)$ was constrained
mainly by the magnetic susceptibility at $eB=0$
~\cite{Rougemont:2015oea}. It was not fitted directly to finite-field
thermodynamics. Lattice entropy data at nonzero $eB$ provide such a constraint
~\cite{Bali:2014kia}. The magnetic CEP can therefore be determined with
$f(\phi)$ calibrated directly at finite field.

We use the radial holographic equations to calibrate the model against
lattice thermodynamics. The AdS/deep-learning correspondence identifies
radial evolution with the layers of a neural network. Deep learning has
also become a standard tool in the physical sciences
~\cite{LeCun:2015pmr,Carleo:2019ptp}. It has been used to infer holographic
model parameters from QCD, transport, and lattice data
~\cite{Hashimoto:2018ftp,Akutagawa:2020yeo,Hashimoto:2019bih,
Hashimoto:2018bnb,Yan:2020wcd,Chen:2024ckb,Chen:2025goz,Zhu:2025edv,
Zhu:2026dmc}. Neural ordinary differential equations have also been
applied to magnetic and dense QCD phase diagrams~\cite{Cai:2024eqa}.
Here the EMD equations determine the radial layer map, and the finite
sets of potential and coupling parameters are calibrated with
adaptive Nelder--Mead minimisation.

The zero-field background is calibrated to the $eB=0$ entropy of
ref.~\cite{Bali:2014kia}, with the scalar ultraviolet mass held fixed.
WB and HotQCD entropy and sound-speed data
~\cite{Borsanyi:2013bia,HotQCD:2014kol} provide thermodynamic comparisons
rather than additional fitting targets.

The finite-field calibration uses the same zero-field background.
Entropy data at four nonzero fields determine a four-parameter
$f(\phi)$, and three other fields are reserved for comparison.
The coupling remains positive, with monotonic decrease imposed as a model assumption.
The model is then used to calculate entropy, specific heat, pressure,
energy density, trace anomaly, and squared speed of sound.
At fixed physical $eB$, the merger of the two temperature turning points
locates the CEP; the divergence of $C_V$ and the condition $c_s^2=0$
give the corresponding thermodynamic signatures. Above the CEP, entropy
develops an $S$-shaped structure and pressure develops a swallowtail.
The crossover temperature is defined by the transition-region inflection
of $s/T^3$ with the largest positive slope. The Maxwell construction
determines the first-order transition temperature $T_t$ on the same
connected solution family. We find
$(eB_c,T_c)\simeq(7.2200\,\mathrm{GeV}^{2},95.855\,\mathrm{MeV})$.
The critical field lies within the lattice interval of
ref.~\cite{DElia:2021yvk}.

Section~\ref{sec:model} gives the EMD model and its thermodynamic dictionary.
Section~\ref{sec:ML} describes the zero-field and finite-field
calibrations, and
the fixed-$eB$ input map. The numerical method is outlined in
appendix~\ref{app:architecture}. Section~\ref{sec:results} presents the
thermodynamic phase structure and the magnetic CEP. Section~\ref{sec:summary}
summarises the results.

\section{The anisotropic Einstein--Maxwell--dilaton model}
\label{sec:model}

We use the five-dimensional Einstein--Maxwell--dilaton (EMD) action in an
asymptotically anti--de Sitter ($\mathrm{AdS}_5$) spacetime
~\cite{DeWolfe:2010he,Rougemont:2015oea}:
\begin{widetext}
\begin{equation}
\label{eq:action}
S = \frac{1}{2\kappa^2} \int_{M_5} d^5x \sqrt{-g}
      \left[ R - \frac{1}{2}(\partial_\mu \phi)^2 - V(\phi)
             - \frac{f(\phi)}{4} F_{\mu\nu}F^{\mu\nu} \right],
\end{equation}
\end{widetext}
where $\kappa^2=8\pi G_5$. The dilaton $\phi$ introduces the infrared
nonconformal scale~\cite{Gubser:2008ny,Gubser:2008yx}. The Abelian field
$A_\mu$ sources a uniform magnetic field along the $z$ direction. Its
dilaton-dependent coupling is denoted by $f(\phi)$
~\cite{DeWolfe:2010he,Rougemont:2015oea}.

A uniform magnetic field preserves rotations in the transverse plane but
breaks the spatial $SO(3)$ symmetry~\cite{DHoker:2009mmn}. We adopt the
anisotropic black-brane ansatz of ref.~\cite{Rougemont:2015oea},
\begin{equation}
\label{eq:metric}
\begin{aligned}
ds^2 ={}& e^{2a(r)}\!\left[-h(r)dt^2 + dz^2\right]\\
 &+ e^{2c(r)}\!\left(dx^2 + dy^2\right)
 + \frac{e^{2b(r)}}{h(r)}dr^2\,.
\end{aligned}
\end{equation}
\begin{equation}
\phi = \phi(r)\,, \qquad
A = \mathcal{B}\,x\,dy \;\Rightarrow\; F = dA = \mathcal{B}\,dx\wedge dy\,,
\end{equation}
where $r$ is the radial coordinate, and the conformal boundary lies at
$r\to\infty$. The horizon $r_H$ is the largest root of $h(r_H)=0$. In the
numerical coordinates, the dimensionless magnetic component is
$\mathcal B\equiv F_{xy}$. Section~\ref{sec:dict} relates it to the
standard-coordinate component $\tilde{\mathcal B}$ and to the physical field
$eB$. We use the domain-wall gauge $b(r)=0$.

\subsection{Equations of motion and model functions}
\label{sec:action}

Substitution of the ansatz into eq.~\eqref{eq:action} gives four second-order
equations~\cite{Rougemont:2015oea}:
\begin{widetext}
\begin{align}
\phi'' + \left(2a' + 2c' + \frac{h'}{h}\right)\phi'
   &- \frac{1}{h}\!\left(\frac{\partial V}{\partial\phi}
     + \frac{\mathcal{B}^2 e^{-4c}}{2}\frac{\partial f}{\partial\phi}\right) = 0\,,\\[2pt]
a'' + \left(\frac{14}{3}c' + \frac{4}{3}\frac{h'}{h}\right)a'
   &+ \frac{8}{3}a'^2 + \frac{2}{3}c'^2 + \frac{2}{3}\frac{h'}{h}c'
   + \frac{2}{3h}V(\phi) - \frac{1}{6}\phi'^2 = 0\,,\\[2pt]
c'' - \left(\frac{10}{3}a' + \frac{1}{3}\frac{h'}{h}\right)c'
   &+ \frac{2}{3}c'^2 - \frac{4}{3}a'^2 - \frac{2}{3}\frac{h'}{h}a'
   - \frac{1}{3h}V(\phi) + \frac{1}{3}\phi'^2 = 0\,,\\[2pt]
h'' + \left(2a' + 2c'\right)h' &= 0\,.
\end{align}
They are supplemented by the first-order Hamiltonian constraint
\begin{equation}
a'^2 + c'^2 - \frac{1}{4}\phi'^2
  + \left(\frac{a'}{2} + c'\right)\frac{h'}{h} + 4a'c'
  + \frac{1}{2h}\!\left[V(\phi) + \frac{\mathcal{B}^2 e^{-4c}}{2}f(\phi)\right] = 0\,.
\end{equation}
\end{widetext}
This constraint is imposed together with the second-order equations and
ensures consistency of the resulting gravitational solution.

We take a cosh-plus-even-polynomial form for the dilaton potential. The
cosh-plus-quadratic form was introduced in ref.~\cite{Gubser:2008ny}.
Quartic and sextic terms were later used in QCD-calibrated holographic models
~\cite{Finazzo:2014cna,Rougemont:2015oea,Critelli:2017oub,Cao:2024jgt}:
\begin{equation}
\label{eq:V_ansatz}
V(\phi) = -12\cosh(v_1 \phi) + v_2 \phi^2 + v_3 \phi^4 + v_4 \phi^6\,.
\end{equation}
We set $L=1$, so $V(0)=-12$. Near $\phi=0$, the potential is
$V(\phi)=-12+\frac12m^2\phi^2+\cdots$, where
\begin{equation}
\label{eq:m2_potential}
m^2=2\left(v_2-6v_1^2\right)\,.
\end{equation}

The updated zero-field entropy fit determines five independent
parameters $(v_1,v_3,v_4,\kappa^2,\Lambda)$. The fixed ultraviolet mass
determines $v_2$ through eq.~\eqref{eq:V_UV_relation}, as described in
section~\ref{sec:V_calib}. For the Maxwell--dilaton coupling, we use
\begin{equation}
\label{eq:f_ansatz}
\begin{aligned}
f(\phi)&=F_0\bigl[u+c_2u(1-u)+c_3u(1-u)(2u-1)\bigr],\\
u&=\mathrm{sech}(k\phi^2)\,.
\end{aligned}
\end{equation}
The four parameters $(F_0,k,c_2,c_3)$ are calibrated in
section~\ref{sec:f_calib}, including the overall normalisation $F_0$.
The use of a $\mathrm{sech}$-type coupling is motivated by earlier
magnetised EMD models~\cite{Rougemont:2015oea,Finazzo:2016mhm}.
The two polynomial corrections in $u$ vary its shape while preserving
$f(0)=F_0$ and decay as $\phi\to\infty$. We impose $F_0>0$, $k>0$,
and the conditions $f(\phi)>0$ and $f'(\phi)\leq0$ on the complete
function. The coefficients $c_2$ and $c_3$ are not individually
restricted in sign. Positivity gives the Maxwell field positive
kinetic energy~\cite{Rougemont:2015oea}. Given the limited finite-field
thermodynamic data, monotonic decrease is imposed as a modelling assumption.
The continuous shape check is described in appendix~\ref{app:loss-f}.

\subsection{Near-horizon and ultraviolet expansions}
\label{sec:boundaries}

We impose regularity at the horizon and match the numerical solution to its
ultraviolet asymptotics. Near $r=r_H$, the fields are expanded as
~\cite{Rougemont:2015oea,Cao:2024jgt}
\begin{equation}
X(r)=\sum_{n=0}^{\infty} X_n(r-r_H)^n\,, \qquad X\in\{a,c,h,\phi\}\,.
\end{equation}
The second-order expansion contains twelve coefficients. We place the horizon
at $r_H=0$ by a radial translation. Regularity gives $h_0=h(r_H)=0$. The
remaining coordinate freedom is fixed by $a_0=c_0=0$ and
$h_1=h'(r_H)=1$:
\begin{equation}
\label{eq:horizon_gauge}
r_H=0\,,\qquad a_0=c_0=h_0=0\,,\qquad h_1=1\,.
\end{equation}
The independent horizon data are $\phi_0\equiv\phi(r_H)$ and $\mathcal B$.
Substitution into the equations fixes
$(a_1,a_2,c_1,c_2,h_2,\phi_1,\phi_2)$ algebraically. The numerical integration
starts just outside the horizon at a small positive $r=\varepsilon$. Initial values are
obtained from the truncated horizon expansion. The relevant coefficients are summarised in
appendix~\ref{app:initial-layer}.

Near the boundary, the fields approach $\mathrm{AdS}_5$ with the asymptotic
forms~\cite{Rougemont:2015oea,Cao:2024jgt}
\begin{widetext}
\begin{equation}
\label{eq:far-region-expansion}
\begin{aligned}
a(r) &\approx a_0^{\mathrm{far}} + \frac{r}{\sqrt{h_0^{\mathrm{far}}}}\,,
&\qquad
c(r) &\approx c_0^{\mathrm{far}} + \frac{r}{\sqrt{h_0^{\mathrm{far}}}}\,,
\\
h(r) &\approx h_0^{\mathrm{far}} + h_4^{\mathrm{far}}\,e^{-4\alpha(r)} + \cdots\,,
&\qquad
\phi(r) &\approx \phi_A\,e^{-\nu\alpha(r)} + \phi_B\,e^{-\Delta\alpha(r)} + \cdots\,.
\end{aligned}
\end{equation}
\end{widetext}
where $\alpha(r)=a_0^{\mathrm{far}}+r/\sqrt{h_0^{\mathrm{far}}}$. In standard
quantisation, $\phi_A$ and $\phi_B$ are the source and response coefficients
of the dual scalar operator. Its scaling dimension $\Delta$ is related to the
bulk mass by~\cite{Klebanov:1999tb}
\begin{equation}
m^2 L^2 = \Delta(\Delta-4)\,, \qquad \nu \equiv 4-\Delta\,.
\end{equation}
The updated background fixes $\nu=1.26706$, $\Delta=2.73294$, and
$m^2=-\nu\Delta$. The potential and ultraviolet fits at zero and finite field use
these same exponents. The mass satisfies the Breitenlohner--Freedman
bound~\cite{Breitenlohner:1982bm,Breitenlohner:1982jf}.

We extract $a_0^{\mathrm{far}}$, $c_0^{\mathrm{far}}$,
$h_0^{\mathrm{far}}$, and $\phi_A$ by fitting the numerical solution in the
asymptotic region. The extraction method and numerical checks are described in
appendix~\ref{app:hyperparameters}.
For example, figure~\ref{fig:zero_radial_profiles} shows the zero-field
profiles at $T=160\,\mathrm{MeV}$ as they approach the AdS asymptotics.
For this solution, the leading ultraviolet source and temperature remain
stable when the fitting window is varied.

\subsection{Holographic dictionary and thermodynamic observables}
\label{sec:dict}

In the numerical coordinates, $h(r)$ approaches the non-unit constant
$h_0^{\mathrm{far}}$ at the boundary. The usual holographic expressions for
the thermodynamic variables cannot yet be applied. We introduce standard
coordinates in the same domain-wall gauge. The metric keeps the form of
eq.~\eqref{eq:metric}, and its ultraviolet behaviour is normalised as
\begin{equation}
\tilde a(\tilde r)\approx \tilde r\,,\qquad
\tilde c(\tilde r)\approx \tilde r\,,\qquad
\tilde h(\tilde r)\approx 1\,,\qquad
\tilde\phi(\tilde r)\approx e^{-\nu \tilde r}\,.
\end{equation}
The same two-form has the following components in the two coordinate systems:
\begin{equation}
F=\mathcal B\,dx\wedge dy
 =\tilde{\mathcal B}\,d\tilde x\wedge d\tilde y\,.
\end{equation}
Equality of the line elements and two-forms, together with
$\tilde\phi(\tilde r)=\phi(r)$, gives
~\cite{Rougemont:2015oea,Cao:2024jgt}
\begin{widetext}
\begin{subequations}
\label{eq:coordmap}
\begin{align}
\tilde r &= \frac{r}{\sqrt{h_0^{\mathrm{far}}}}
+a_0^{\mathrm{far}}-\ln\!\left(\phi_A^{1/\nu}\right),
&
\tilde t &= \phi_A^{1/\nu}\sqrt{h_0^{\mathrm{far}}}\,t,
\label{eq:coordmap-rt}
\\
\tilde x &= \phi_A^{1/\nu}
e^{\,c_0^{\mathrm{far}}-a_0^{\mathrm{far}}}\,x,
&
\tilde y &= \phi_A^{1/\nu}
e^{\,c_0^{\mathrm{far}}-a_0^{\mathrm{far}}}\,y,
\label{eq:coordmap-xy}
\\
\tilde z &= \phi_A^{1/\nu}\,z,
&
\tilde{\mathcal B} &= \frac{e^{2(a_0^{\mathrm{far}}-c_0^{\mathrm{far}})}}
{\phi_A^{2/\nu}}\,\mathcal B,
\label{eq:coordmap-zB}
\\[4pt]
\tilde a(\tilde r) &= a(r)-\ln\!\left(\phi_A^{1/\nu}\right),
&
\tilde c(\tilde r) &= c(r)-\left(c_0^{\mathrm{far}}-a_0^{\mathrm{far}}\right)
-\ln\!\left(\phi_A^{1/\nu}\right),
\label{eq:coordmap-ac}
\\
\tilde h(\tilde r) &= \frac{h(r)}{h_0^{\mathrm{far}}},
&
\tilde\phi(\tilde r) &= \phi(r).
\label{eq:coordmap-hphi}
\end{align}
\end{subequations}
\end{widetext}
Tildes distinguish the standard coordinates from the numerical coordinates.
Thus $\mathcal B$ and $\tilde{\mathcal B}$ are components of the same bulk
two-form in different coordinates. Physical boundary quantities are written
as $(T,eB,s)$ without tildes. We find $\phi_A>0$ for every solution used
below. The fractional powers and logarithm in eq.~\eqref{eq:coordmap} are
therefore real.

The plasma temperature is the Hawking temperature of the black-brane horizon.
The entropy density follows from the Bekenstein--Hawking area law
~\cite{Bekenstein:1973ur,Hawking:1975vcx}. Using eq.~\eqref{eq:coordmap}, the
horizon normalisation, and the scale $\Lambda$, we obtain
~\cite{Rougemont:2015oea,Cao:2024jgt}
\begin{equation}
\label{eq:physical_observables}
\begin{aligned}
T &= \frac{\Lambda}
{4\pi\,\phi_A^{1/\nu}\sqrt{h_0^{\mathrm{far}}}}\,,\\[2pt]
eB &= \frac{\Lambda^2 e^{2(a_0^{\mathrm{far}}-c_0^{\mathrm{far}})}}
{\phi_A^{2/\nu}}\,\mathcal B\,,\\[2pt]
s &= \frac{2\pi\Lambda^3 e^{2(a_0^{\mathrm{far}}-c_0^{\mathrm{far}})}}
{\kappa^2\phi_A^{3/\nu}}\,.
\end{aligned}
\end{equation}
The scale $\Lambda$ restores physical units. The electromagnetic coupling is
included in the conventional boundary quantity $eB$.

At fixed $eB$, entropy determines pressure differences, leaving an
additive pressure constant. The finite-field calculation chooses a
common reference temperature $T_{\rm low}$ and sets
$p(T_{\rm low},eB)=0$ on its slices. This thermodynamic-integration
prescription is used in magnetised EMD models
~\cite{Rougemont:2015oea,Cao:2024jgt}:
\begin{equation}
\label{eq:thermo_derived}
p(T,eB) \;=\; \int_{T_{\rm low}}^{T} s(T',eB)\,dT'\,,
\qquad
\varepsilon(T,eB) \;=\; Ts-p\,.
\end{equation}
For a multivalued solution curve, the integral follows the parametrised
solution curve at fixed $eB$, using one common reference normalisation.
For the updated pure zero-field diagnostics, we instead anchor the pressure
at $T_*=115\,\mathrm{MeV}$ using $p(T_*,0)/T_*^4=0.274$ from
ref.~\cite{Borsanyi:2013bia}, and use
\begin{equation}
\label{eq:zero_pressure_anchor}
p(T,0)=p(T_*,0)+\int_{T_*}^{T}s(u,0)\,du.
\end{equation}
This anchor is not fitted by the entropy objective. Energy density and
trace anomaly for the updated background use this same anchored pressure.
\begin{equation}
\label{eq:thermo_response}
C_{V,eB} \;=\; T\!\left(\frac{\partial s}{\partial T}\right)_{\!eB}\,,
\qquad
c_{s,eB}^{\,2} \;\equiv\;\left(\frac{\partial p}{\partial\varepsilon}\right)_{\!eB}
 = \frac{s}{C_{V,eB}}\,.
\end{equation}
We write $c_s^2$ and $C_V$ when the fixed-$eB$ condition is clear. All
derived quantities within each calculation use its specified pressure
normalisation. We adopt the fixed-field
($B$-scheme) convention. In this scheme, $p$ is isotropic and enters
$\varepsilon=Ts-p$. In the fixed-flux ($\Phi$-scheme) convention, the
longitudinal pressure equals the $B$-scheme pressure. The transverse pressure
also contains a magnetisation term~\cite{Bali:2013esa,Bali:2014kia}. We define
the fixed-field trace anomaly by\footnote{At $eB=0$, $I$ is the usual
isotropic trace anomaly $\Theta=\varepsilon-3p$. In the $B$-scheme,
$p_\parallel=p_\perp=p$ and
$I^{(B)}=\varepsilon-p_\parallel-2p_\perp=\varepsilon-3p$. In the
$\Phi$-scheme, $p_\parallel=p$ and
$p_\perp=p-eB\,\mathcal M$, where $\mathcal M$ is the magnetisation. Hence
$I^{(\Phi)}=I^{(B)}+2eB\,\mathcal M$
~\cite{Bali:2013esa,Bali:2014kia}.}
\begin{equation}
\label{eq:trace_anomaly}
I(T,eB)\equiv\varepsilon(T,eB)-3p(T,eB)\,.
\end{equation}
Equations~\eqref{eq:coordmap} and~\eqref{eq:physical_observables} map
$(\phi_0,\mathcal B)$ to $(T,eB)$. The other thermodynamic quantities follow
from eqs.~\eqref{eq:thermo_derived}--\eqref{eq:trace_anomaly}.
Section~\ref{sec:ML} uses this map in the model calibration.

\section{Calibration of the EMD model}
\label{sec:ML}

\subsection{Calibration strategy}
\label{sec:calibration_strategy}

We calibrate the finite sets of coefficients in
eqs.~\eqref{eq:V_ansatz} and~\eqref{eq:f_ansatz}. The radial EMD equations
are solved by a fourth-order Runge--Kutta (RK4) map. Each radial layer
performs one RK4 step, with the bulk fields and their derivatives as
its state. The zero-field profiles in
figure~\ref{fig:zero_radial_profiles} illustrate this evolution.
The dictionary in section~\ref{sec:dict} converts the ultraviolet
output to physical observables. This radial-layer interpretation follows
the AdS/deep-learning construction of
refs.~\cite{Hashimoto:2018ftp,Hashimoto:2018bnb,Hashimoto:2019bih,
Akutagawa:2020yeo}. Related data-driven holographic studies are given in
refs.~\cite{Zhu:2025edv,Zhu:2026dmc,Chen:2024ckb,Chen:2025goz}.

The zero-field fit uses bounded adaptive Nelder--Mead to determine
$(v_1,v_3,v_4,\kappa^2,\Lambda)$, while $v_2$ follows from the fixed
ultraviolet mass. The finite-field fit uses adaptive Nelder--Mead to
determine $(F_0,k,c_2,c_3)$ with that background held fixed.
Figure~\ref{fig:DNN-architecture} shows the two calculations.
The numerical method and loss functions are given in appendix~\ref{app:architecture}.

\begin{figure*}[tp]
\centering
\begin{tikzpicture}[font=\scriptsize, >=Latex,
  box/.style={draw, rounded corners=2pt, align=center, minimum height=8mm,
              text width=28mm, inner sep=2pt},
  solver/.style={box, fill=blue!8, text width=36mm},
  input/.style={box, fill=green!9},
  data/.style={box, fill=yellow!14},
  loss/.style={box, fill=orange!13},
  freeze/.style={draw, rounded corners=2pt, fill=gray!14, align=center,
                 inner sep=2pt, text width=32mm}]
\node[input] (zeroinput) {horizon input\\$\phi_0$};
\node[solver, right=8mm of zeroinput] (zerosolver) {Zero-field EMD solver\\RK4 evolution $\to$ UV dictionary};
\node[loss, right=8mm of zerosolver] (zeroloss) {$\mathcal L_0[s/T^3]$\\update $v_1,v_3,v_4,\kappa^2,\Lambda$};
\node[data, above=6mm of zeroloss] (zerodata) {Bali $eB=0$\\entropy data};
\draw[->] (zeroinput) -- (zerosolver);
\draw[->] (zerosolver) -- (zeroloss);
\draw[->] (zerodata) -- (zeroloss);
\node[freeze, below=8mm of zeroloss] (frozen) {Bali zero-field background\\Table~\ref{tab:V_params_20261005}};
\node[input, below=22mm of zeroinput] (fieldinput) {horizon inputs\\$(\phi_0,\mathcal B)$};
\node[solver, right=8mm of fieldinput] (fieldsolver) {Finite-field EMD solver\\with frozen $V(\phi)$; fixed-$eB$ roots};
\node[loss, right=8mm of fieldsolver] (fieldloss) {$\mathcal L_B[s/T^3]$\\update $F_0,k,c_2,c_3$};
\node[data, below=6mm of fieldloss] (fielddata) {finite-field entropy\\lattice targets};
\draw[->] (fieldinput) -- (fieldsolver);
\draw[->] (fieldsolver) -- (fieldloss);
\draw[->] (fielddata) -- (fieldloss);
\draw[->] (frozen.south) -- ++(0,-2mm) -| (fieldsolver.north);
\end{tikzpicture}
\caption{Calibration of the zero-field background (upper) and the
Maxwell--dilaton coupling (lower). The zero-field fit uses the Bali
entropy data and derivative-free minimisation. The finite-field fit
uses the same background in Table~\ref{tab:V_params_20261005}
and the same weighted-band objective strategy.}
\label{fig:DNN-architecture}
\end{figure*}
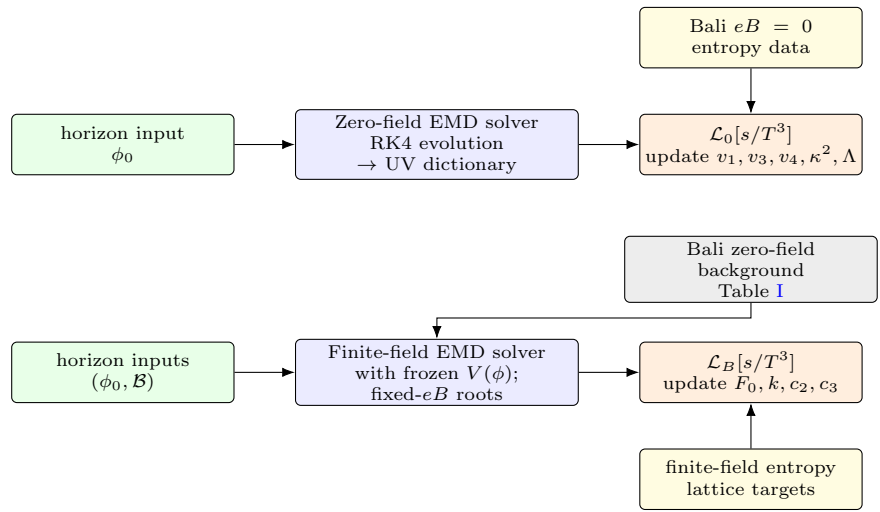

\subsection{Zero-field calibration of the dilaton sector}
\label{sec:V_calib}

At $eB=0$, the Maxwell term vanishes and the geometry is isotropic,
$a(r)=c(r)=A(r)$. We recalibrate this background to the zero-field
entropy column of the ancillary equation-of-state table in
ref.~\cite{Bali:2014kia} over its temperature range,
$114$--$305\,\mathrm{MeV}$. We fit $\mathcal S=s/T^3$ using the
reported entropy bands. This zero-field slice gives the background
the same data baseline as the finite-field calibration in
section~\ref{sec:f_calib}.

The potential has the form in eq.~\eqref{eq:V_ansatz}. Its five
independent parameters are
$(v_1,v_3,v_4,\kappa^2,\Lambda)$. We fix the ultraviolet exponents
$\nu=1.26706$ and $\Delta=2.73294$, and derive
\begin{equation}
\label{eq:V_UV_relation}
v_2=6v_1^2-\frac{\nu\Delta}{2}.
\end{equation}
Consequently, $m^2=-\nu\Delta$ remains fixed throughout the
calibration. The scalar mass and the ultraviolet fit use the same
exponents. The parameter domain is $v_1>0$, $v_2\geq0$, $v_3\leq0$,
$v_4\geq0$, $\kappa^2>0$, and $\Lambda>0$.

The radial equations are solved with RK4, and a natural cubic
interpolant of the model entropy is evaluated at the lattice
temperatures. Accepted solutions must cover the reference range;
no extrapolation is used in the fit.

Only entropy enters the weighted-band loss in
appendix~\ref{app:loss-V}. The parameters are optimized with bounded,
scaled adaptive Nelder--Mead, without using backpropagated gradients.
We start from a previously fitted parameter set and refine it locally.
Since the earlier fits also used sound-speed data, the sound-speed
comparisons below are not independent validation tests.
Table~\ref{tab:V_params_20261005} gives the resulting background.

\begin{table*}[tp]
\centering
\small
\setlength{\tabcolsep}{4pt}
\renewcommand{\arraystretch}{1.3}
\begin{tabular}{cccccc}
\hline\hline
$v_1$ & $v_2$ & $v_3$ & $v_4$ & $\kappa^2$ & $\Lambda\;(\mathrm{MeV})$\\
\hline
$0.5826099273$ & $0.3052064859$ & $-0.04645832525$ &
$0.000785410036$ & $11.13878128$ & $1155.390747$\\
\hline\hline
\end{tabular}
\caption{Recalibrated background obtained from the $eB=0$ entropy
column of ref.~\cite{Bali:2014kia} over $114$--$305\,\mathrm{MeV}$.
There are five independent parameters; $v_2$ follows from
eq.~\eqref{eq:V_UV_relation}. Here $\Lambda$ is the energy scale.}
\label{tab:V_params_20261005}
\end{table*}

Independent calculations at increased resolution check the fitted
thermodynamics. The calibration satisfies local convergence checks,
without establishing a global minimum or a unique parameter set.

\begin{figure*}[tp]
\centering
\includegraphics[width=\textwidth]{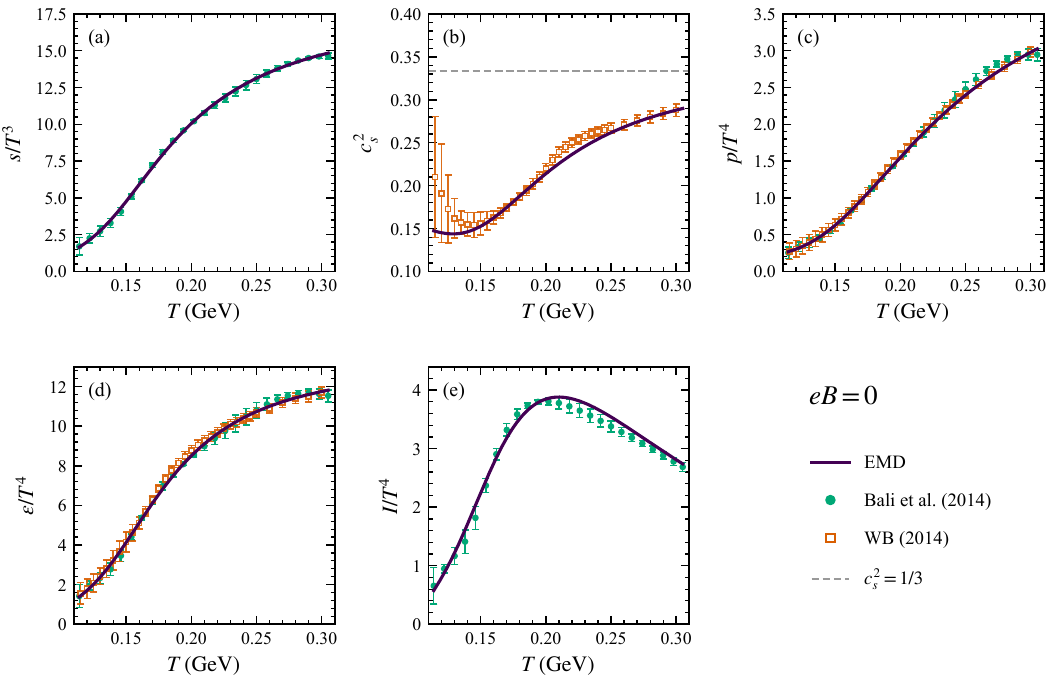}
\caption{Zero-field calibration and thermodynamic comparisons: (a) $s/T^3$, (b) $c_s^2$, (c) $p/T^4$,
(d) $\varepsilon/T^4$, and (e) $I/T^4$.
Green circles show the Bali table~\cite{Bali:2014kia}; open orange
squares show the WB table~\cite{Borsanyi:2013bia} in panels (b)--(d).
Only the Bali entropy enters the objective; panel (a) is
therefore a fit comparison. Sound speed is derived from that entropy.
Pressure uses the fixed WB anchor at $115\,\mathrm{MeV}$;
energy density and trace anomaly follow from the same entropy and pressure.
Panels (b)--(e) are conditional diagnostics, not independent fits.
Error bars show the uncertainties reported in the respective
lattice tables.}
\label{fig:zero_magnetic_field}
\end{figure*}

Figure~\ref{fig:zero_magnetic_field} collects the zero-field fit and
thermodynamic comparisons for the common background in
table~\ref{tab:V_params_20261005}.
Writing $\mathcal S=s/T^3$, thermodynamic consistency gives
\begin{equation}
\label{eq:cs_from_entropy_spline}
c_s^2=\frac{\mathcal S}{3\mathcal S+T\mathcal S'}.
\end{equation}
Accepted entropy curves obey positivity, thermodynamic stability,
and $0<c_s^2\leq1$ over the fitted interval. The entropy is largely
compatible with its reported band, while the derived sound speed and
other equation-of-state observables retain deviations from the lattice
central values. These comparisons are diagnostics rather than
independent fitting constraints.

Pressure differences follow from the same entropy curve. For the
zero-field pressure diagnostic, the additive constant is fixed by
the WB central value $p/T^4=0.274$ at $115\,\mathrm{MeV}$.
Entropy alone does not determine this constant, so the pressure
comparison depends on the chosen anchor. It is a consistency check,
since the anchor and the comparison data come from the same
equation-of-state table.

This fit gives a deterministic point estimate. The objective uses the
reported bands without assuming independent Gaussian errors across
temperatures; it does not yield a Bayesian posterior.
The resulting background in table~\ref{tab:V_params_20261005} is held
fixed throughout the finite-field coupling calibration.

\subsection{Finite-field calibration and validation of the Maxwell--dilaton sector}
\label{sec:f_calib}

At finite field, only the four parameters of $f(\phi)$ are trained.
This function controls the magnetic response
~\cite{Rougemont:2015oea,Cao:2024jgt}. The radial solver evolves
backgrounds labelled by $(\phi_0,\mathcal B)$, while the zero-field
parameters remain fixed at the values in
table~\ref{tab:V_params_20261005}.

The entropy data at $eB=0.1$, $0.2$, $0.3$, and
$0.4\,\mathrm{GeV}^2$~\cite{Bali:2014kia} determine the coupling.
The fields $eB=0.5$, $0.6$, and $0.7\,\mathrm{GeV}^2$ are withheld
for comparison. The zero-field data determine the background and are
not an independent test.

All training fields use the same weighted-band entropy objective,
defined in appendix~\ref{app:loss-f}. Magnetic susceptibility is not
included. The parameters are determined by adaptive Nelder--Mead,
subject to positivity and monotonicity of the complete coupling.
Table~\ref{tab:f_params} gives the resulting parameters.

\begin{table*}[tp]
\centering
\renewcommand{\arraystretch}{1.3}
\begin{tabular}{cccc}
\hline\hline
$F_0$ & $k$ & $c_2$ & $c_3$ \\
\hline
$0.9671145952$ & $0.1015403735$ &
$1.001232414\times10^{-6}$ & $0.9999988660$ \\
\hline\hline
\end{tabular}
\caption{Parameters of the coupling in eq.~\eqref{eq:f_ansatz}, fitted
to the full entropy data at $eB=0.1$, $0.2$, $0.3$, and
$0.4\,\mathrm{GeV}^2$~\cite{Bali:2014kia}. The background in
table~\ref{tab:V_params_20261005} is held fixed.}
\label{tab:f_params}
\end{table*}

We restrict the horizon inputs to the regular solution family connected
to $\mathcal B=0$, requiring that each solution reach the ultraviolet
fitting region. The numerically accessible field range depends on the
fitted $V(\phi)$ and $f(\phi)$ and on the integration and acceptance
criteria; it is not a physical upper bound or a perturbative stability
boundary.

\begin{figure}[tbp]
\centering
\includegraphics[width=0.95\linewidth]{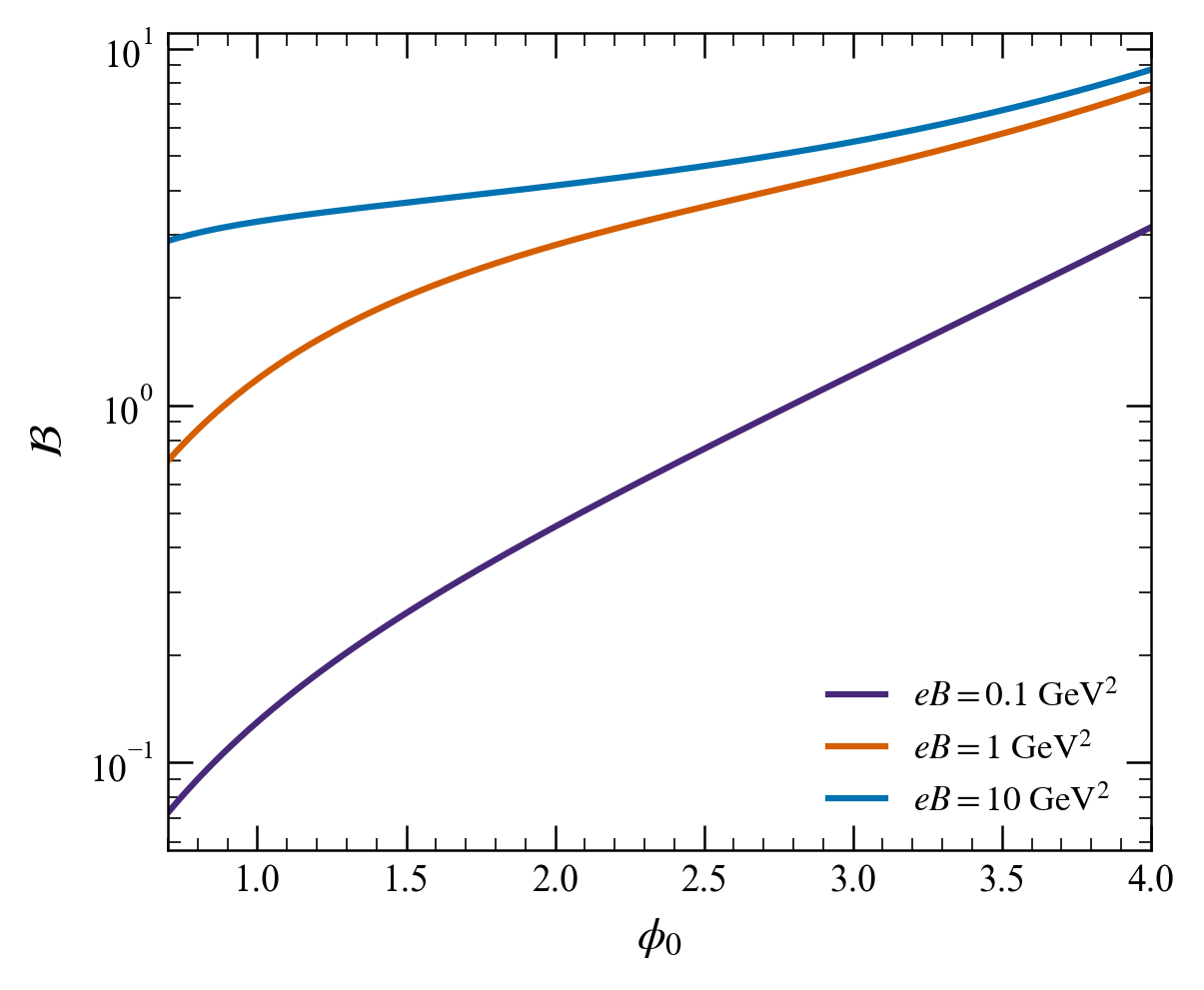}
\caption{Input map for the fixed-field thermodynamic curves.
Each line gives the horizon inputs $(\phi_0,\mathcal B)$ on the
connected solution family at $eB=0.1$, $1$, or
$10\,\mathrm{GeV}^2$. The vertical axis is logarithmic.
The thermodynamic turning points are determined from $T(\phi_0)$
at fixed physical field.}
\label{fig:input_map}
\end{figure}

We next construct thermodynamic curves at fixed physical field.
The dictionary maps $(\phi_0,\mathcal B)$ to $(T,eB)$.
At each $\phi_0$, we solve the equation $eB(\phi_0,\mathcal B)=eB_\star$
on the regular family connected to $\mathcal B=0$.
For training predictions, the two horizon variables are solved together
to cover the prescribed temperature interval at fixed $eB_\star$.
Appendix~\ref{app:architecture} gives the construction.
Figure~\ref{fig:input_map} shows the horizon inputs for
$eB=0.1$, $1$, and $10\,\mathrm{GeV}^2$.
Small $\phi_0$ corresponds to high temperature away from the
strong-field turning-point interval. The relation between the
numerical field $\mathcal B$ and the physical field $eB$ is nonlinear.

Different points on a fixed-field solution family can produce the
same physical temperature. In the strong-field regime this gives the
multivalued thermodynamic curves discussed in section~\ref{sec:results}.
The transition criteria are applied to the thermodynamic curves,
with their connected branch ordering preserved.

\begin{figure}[tbp]
\centering
\includegraphics[width=0.98\linewidth]{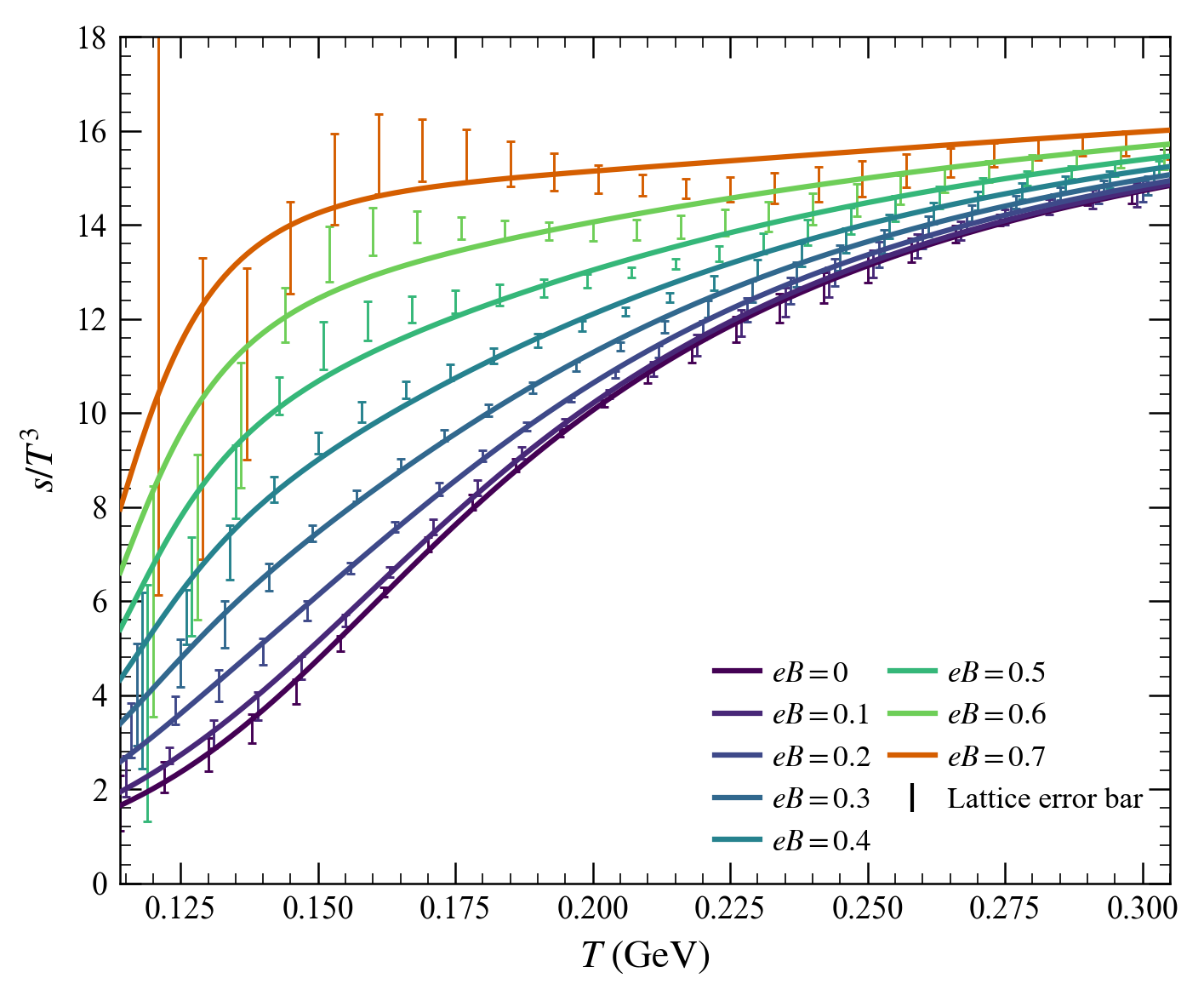}
\caption{Entropy density at the eight lattice fields. Error bars show
selected tabulated entropy data of ref.~\cite{Bali:2014kia}, and
solid lines are the calibrated EMD results with the background in
table~\ref{tab:V_params_20261005}.
The $eB=0$ curve is the fitted background and is not independent
validation. Fields $eB=0.1$, $0.2$, $0.3$, and
$0.4\,\mathrm{GeV}^2$ train the coupling, while $eB=0.5$, $0.6$, and
$0.7\,\mathrm{GeV}^2$ are withheld. The curves are ordered from bottom
to top with increasing $eB$.}
\label{fig:entropy_all_eB}
\end{figure}

We compare the predictions at the three fields excluded from the loss.
Figure~\ref{fig:entropy_all_eB} shows the overall temperature dependence
and field ordering of the entropy data~\cite{Bali:2014kia}.
Residual deviations remain in both the training and withheld curves.
The deviations become appreciable at $eB=0.7\,\mathrm{GeV}^2$.
Thus the withheld fields test a limited extrapolation of the learned
field dependence, rather than establishing agreement across all bands.

\section{Magnetic-field-driven phase structure and the critical end point}
\label{sec:results}

The magnetic results in this section use the background in
table~\ref{tab:V_params_20261005} and the coupling in
Table~\ref{tab:f_params}, including the $eB=0$ curves in mixed-field
figures. Fixed-field curves are constructed by solving for the
physical field as described in section~\ref{sec:ML}.
Entropy distinguishes the crossover, critical, and first-order regimes.
Pressure, energy density, trace anomaly, and squared speed of sound
provide coexistence and local stability checks. The corresponding
transition temperatures form the magnetic phase diagram.
Results outside the lattice training interval are predictions of the
fitted EMD model on the selected connected solution family.

\subsection{Magnetic-field dependence of the entropy density}
\label{sec:criticality}

\begin{figure*}[tp]
\centering
\begin{minipage}[b]{0.48\textwidth}
\centering
\includegraphics[width=\linewidth]{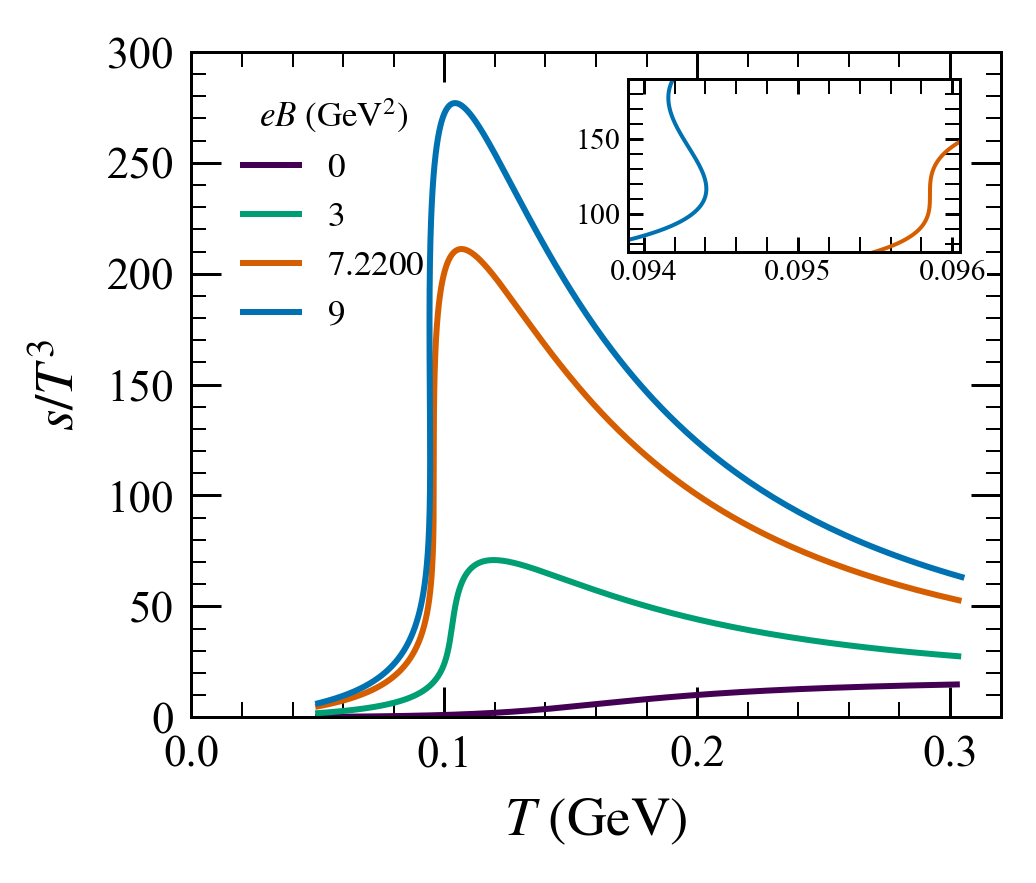}
\end{minipage}\hfill
\begin{minipage}[b]{0.48\textwidth}
\centering
\includegraphics[width=\linewidth]{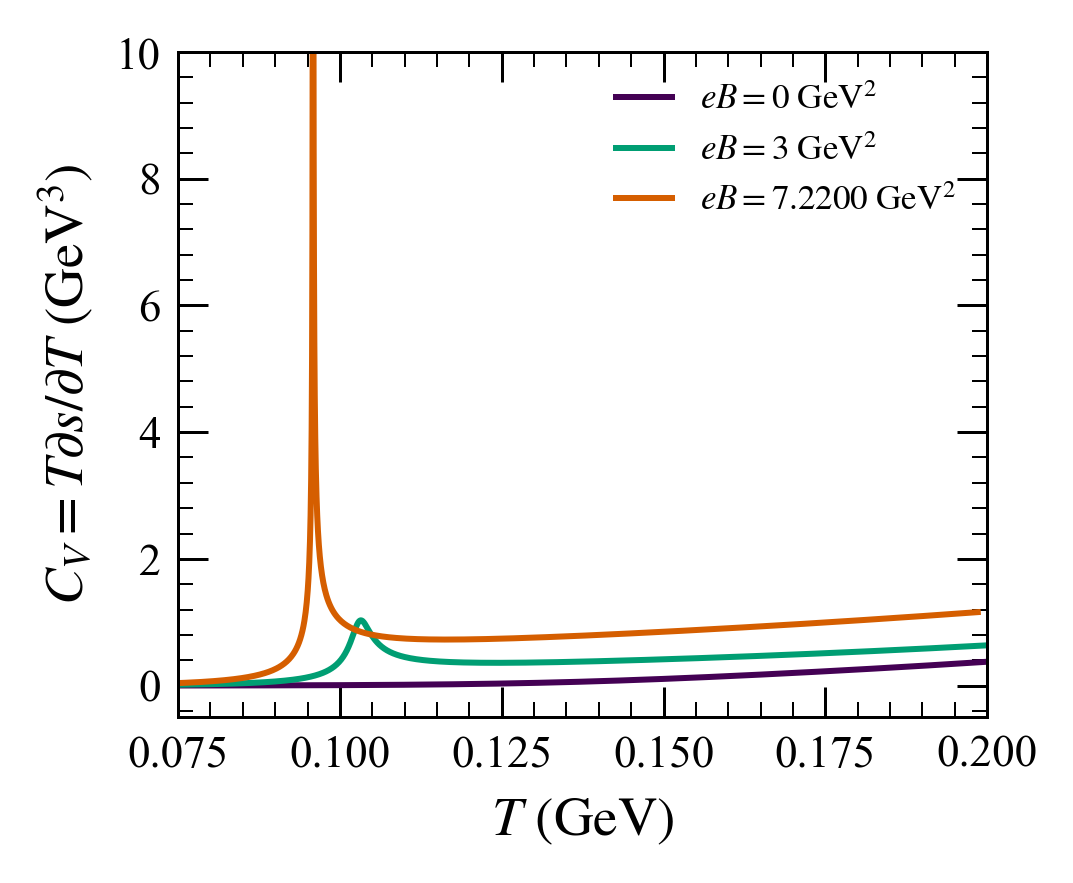}
\end{minipage}
\caption{Entropy density and specific heat near the change of order. Left panel shows $s/T^3$ at $eB=0$, $3$, $7.2200$, and $9\,\mathrm{GeV}^{2}$. The
subcritical curves are smooth and single-valued. At stronger field, a steep segment develops into an $S$-shaped structure. Right panel shows
$C_V=T(\partial s/\partial T)_{eB}$ at $eB=0$, $3$, and
$7.2200\,\mathrm{GeV}^{2}$. The finite-field peak sharpens as the critical field is approached;
at the CEP, the exact response diverges. }
\label{fig:s_cv}
\end{figure*}

We solve the horizon field at fixed physical $eB$ and apply the holographic
dictionary. Figure~\ref{fig:s_cv} shows the resulting entropy curves.

At $eB=0$, $s/T^3$ rises from its low-temperature value and approaches a
nearly conformal plateau at high temperature
~\cite{Borsanyi:2010cj,Borsanyi:2013bia,HotQCD:2014kol,Bali:2014kia}. At
moderate magnetic fields, the entropy is enhanced in the intermediate-
temperature region, consistent with the low-field lattice data
~\cite{Bali:2014kia}. At $eB=3\,\mathrm{GeV}^{2}$, $s/T^3$ rises rapidly through the
crossover region, reaches a maximum, and decreases over the higher-
temperature interval shown. At the critical field the rise becomes
vertical, while the $eB=9\,\mathrm{GeV}^{2}$ curve develops a
multivalued segment. A maximum in $s/T^3$ alone does not establish
a change of transition order. The change from crossover to first order
agrees qualitatively with strong-field lattice and magnetic EMD results
~\cite{Endrodi:2015oba,DElia:2021yvk,Cao:2024jgt}.

The transition sharpens as the critical field is approached.
For subcritical fields, the entropy curve remains single-valued and the
crossover is characterized by a pseudocritical temperature. As $eB$ approaches
the critical value, the pseudocritical temperature approaches the
transition temperature, and the two coincide at the critical end
point. At the same point, the two turning points of the entropy curve merge.
For $eB>eB_c$, the turning points separate and a finite $S$-shaped
structure develops. We locate the critical point directly on the
fixed-field temperature curve and use the specific heat as a
thermodynamic check.

At fixed magnetic field, the specific heat is
\begin{equation}
  C_V=T\left(\frac{\partial s}{\partial T}\right)_{eB} .
\end{equation}
At $eB=0$, $C_V$ is smooth and finite. At $eB=3\,\mathrm{GeV}^{2}$, it has a
rounded peak. The peak sharpens at the limiting field. At the point where the
entropy curve becomes vertical, $(\partial s/\partial T)_{eB}$ diverges.

Smooth entropy and finite $C_V$ identify the low-field crossover. At the CEP,
the two entropy turning points coincide and $C_V$ diverges. Above it, the finite $S$-shaped structure signals a first-order transition. Similar $S$-shaped structures occur in charged holographic thermodynamics~\cite{Chamblin:1999tk,Kubiznak:2012wp} and in finite-density holographic CEP
models~\cite{DeWolfe:2010he,Critelli:2017oub}. Define
$D=(dT/d\phi_0)_{eB}$. At the CEP we solve $D=0$ and
$(dD/d\phi_0)_{eB}=0$, re-solving the physical field at every
finite-difference point. The nonzero third derivative of $T$ and the
nonzero field derivative of $D$ establish the local merger of the
two turning points. The result is
\begin{equation}
  (eB_c,T_c)\simeq
  (7.2200\,\mathrm{GeV}^{2},\,0.0958548\,\mathrm{GeV}) .
  \label{eq:CEP_final}
\end{equation}
The location is stable under derivative-step changes and
increased radial resolution. A common-pressure
construction at nearby supercritical fields confirms local
coexistence approaching this endpoint. Statistical and model
uncertainties are not estimated. The checks concern the selected
connected solution family, without a global free-energy comparison
to unrelated roots. The next subsection examines the resulting
thermodynamic structure.

\subsection{Magnetic-field effects on the equation of state}
\label{sec:coexistence}

The entropy and specific heat show a change in the thermal response. The full
equation of state gives further tests. Pressure determines phase coexistence.
The energy density and trace anomaly describe the magnetic response and check
thermodynamic consistency. The squared speed of sound tests the local thermodynamic stability of the equation of state. We compare $eB=3$, $7.2200$, and $9\,\mathrm{GeV}^{2}$, representing the
crossover, critical, and first-order regimes identified from the entropy
and specific heat.

At fixed $eB$, pressure is obtained by integrating the entropy from the common
reference temperature $T_{\rm low}=0.053\,\mathrm{GeV}$. The same subtraction
constant is used on every branch, as specified in
eq.~\eqref{eq:thermo_derived}
~\cite{Rougemont:2015oea,Bali:2013esa,Bali:2014kia,Cao:2024jgt}. Following
the parametrised black-brane curve gives the pressure along each fixed-field
slice. For $eB>eB_c$, this curve becomes multivalued as a function of the
physical temperature $T$, and the transition temperature
$T_t(eB)$ is determined by equal pressure on the two outer branches:
\begin{equation}
\Delta p(T_t;eB)\equiv
p_{\rm high}(T_t,eB)-p_{\rm low}(T_t,eB)=0 .
\label{eq:maxwell_section4}
\end{equation}
Here ``low'' and ``high'' denote the two outer branches at the same physical
temperature $T_t$. Equation~\eqref{eq:maxwell_section4} is the Maxwell
construction in the pressure representation
~\cite{Chamblin:1999tk,Kubiznak:2012wp,Cao:2024jgt}. In terms of the entropy
representation $T(s;eB)$, the same condition is
\begin{equation}
\int_{s_{\rm low}}^{s_{\rm high}}
\bigl[T(s;eB)-T_t\bigr]\,ds=0 ,
\label{eq:maxwell_area_section4}
\end{equation}
where the integral follows the full multibranch curve. The resulting
$T_t(eB)$ is the transition temperature at which the two phases have equal pressure.

\begin{figure*}[tp]
\centering
\includegraphics[width=0.95\textwidth]{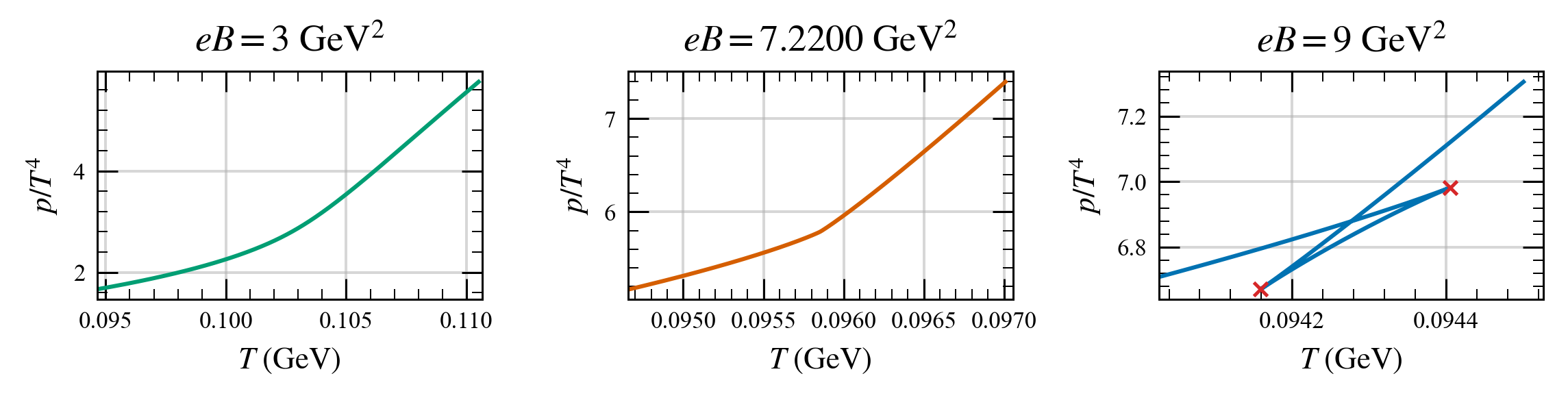}
\caption{Pressure at $eB=3$, $7.2200$, and
$9\,\mathrm{GeV}^{2}$ from left to right, spanning the crossover, critical,
and first-order regimes. Red crosses mark the images of the spinodal
endpoints on the pressure curve. The transition temperature follows from
eq.~\eqref{eq:maxwell_section4}. All curves use the pressure subtraction
in eq.~\eqref{eq:thermo_derived}.}
\label{fig:coexistence}
\end{figure*}

Figure~\ref{fig:coexistence} shows the pressure as a function of the physical
temperature $T$ near the transition at $eB=3$, $7.2200$\, and
$9\,\mathrm{GeV}^{2}$. At the lowest field, the pressure curve is smooth and
single-valued. At the critical field, the swallowtail contracts to a point,
corresponding to the critical temperature
$T_c\simeq0.0958548\,\mathrm{GeV}$. This critical point agrees with the CEP
identified from the entropy and specific heat in section~\ref{sec:criticality}.
At $eB=9\,\mathrm{GeV}^{2}$, a finite swallowtail develops, with its
self-intersection determining the transition temperature $T_t\simeq0.0942768\,\mathrm{GeV}$. The spinodal endpoints delimit the
multibranch interval in the physical temperature $T$, while the
equal-pressure condition determines the transition temperature within this
interval. As $eB\to eB_c^+$, the transition temperature $T_t(eB)$ approaches
the critical temperature $T_c$ obtained in section~\ref{sec:criticality}.

\begin{figure*}[tp]
\centering
\includegraphics[width=0.95\textwidth]{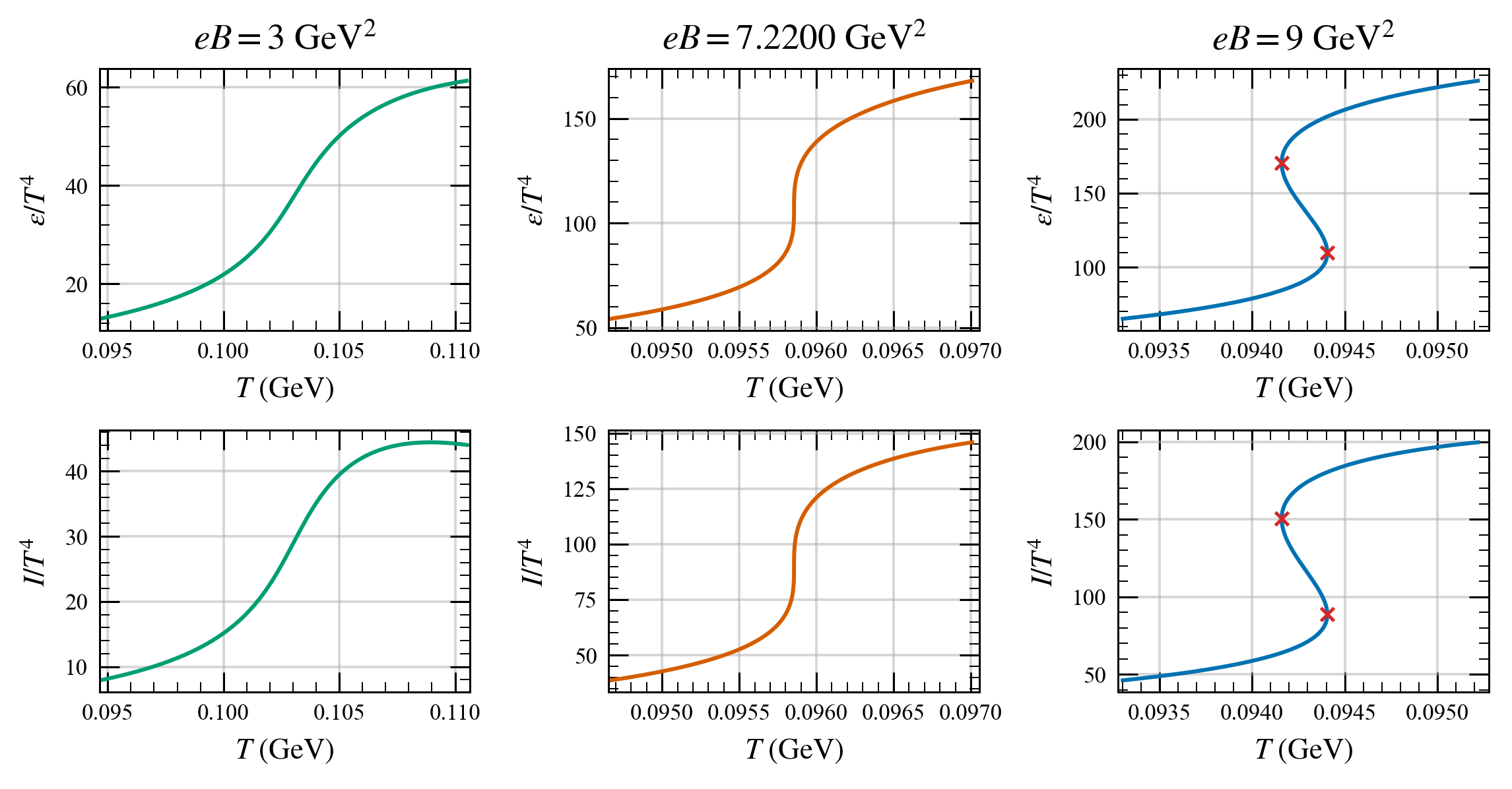}
\caption{Energy density (top) and trace anomaly (bottom) at $eB=3$, $7.2200$,
and $9\,\mathrm{GeV}^{2}$ from left to right. Both quantities are
derived from the same entropy and common-pressure curve and reflect
its corresponding thermodynamic structure.}
\label{fig:energy_trace_anomaly}
\end{figure*}

The pressure swallowtail confirms the coexistence structure inferred from the
entropy. The same $s$ and $p$ give the energy density and trace anomaly through
eqs.~\eqref{eq:thermo_derived} and~\eqref{eq:trace_anomaly}. Their corresponding thermodynamic structure is shown in figure~\ref{fig:energy_trace_anomaly}. These derived quantities provide an additional consistency check of the equation of state and exhibit the same critical structure identified from the entropy and specific heat in section~\ref{sec:criticality}.
At zero field, $I/T^4$ is the usual trace anomaly used to describe conformal
symmetry breaking~\cite{Borsanyi:2013bia,HotQCD:2014kol}. At finite field, we use the $B$-scheme trace anomaly
$I^{(B)}=\epsilon-3p$, where $p=p_z$ is the pressure along the magnetic
field. In the fixed-flux ($\Phi$) scheme, the transverse pressure is
$p_\perp=p_z-eB\,\mathcal M$, and the corresponding trace anomaly is
$I^{(\Phi)}=I^{(B)}+2eB\,\mathcal M$
~\cite{Bali:2013esa,Bali:2014kia}. 

At $eB=3\,\mathrm{GeV}^{2}$, the derived quantities are smooth
and single-valued. Their transition-region variation becomes steeper
at the critical field. At $eB=9\,\mathrm{GeV}^{2}$, they inherit
the multivalued structure of the same connected entropy and pressure
curve. These results depend on the stated common pressure subtraction.

\begin{figure*}[tp]
\centering
\includegraphics[width=0.95\textwidth]{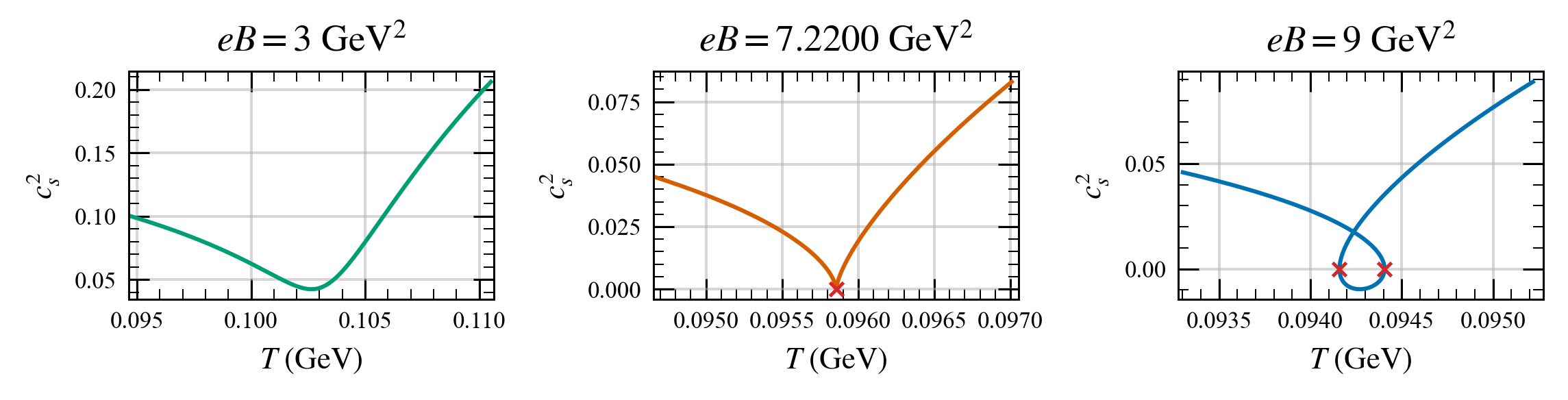}
\caption{Squared speed of sound along fixed-$eB$ slices for
$eB=3$, $7.2200$\, and $9\,\mathrm{GeV}^{2}$, shown from left to right.
At $eB=3\,\mathrm{GeV}^{2}$, below the critical field, $c_s^2$ remains
positive and no spinodal structure is present. At the critical field
$eB_c=7.2200\,\mathrm{GeV}^{2}$, the minimum of $c_s^2$ reaches zero at
$T_c\simeq0.0958548\,\mathrm{GeV}$, corresponding to the merging of the
spinodal turning points. At $eB=9\,\mathrm{GeV}^{2}$, above the critical
field, $c_s^2$ vanishes at the two spinodal endpoints and becomes negative
on the intermediate branch, while remaining positive on the two outer
branches.}
\label{fig:cs2_check}
\end{figure*}

Figure~\ref{fig:cs2_check} shows $c_s^2$ as a function of the physical temperature $T$ along the same fixed-$eB$ slices\, $eB=3$, $7.2200$\, and $9\,\mathrm{GeV}^{2}$. Below the critical field, at $eB=3\,\mathrm{GeV}^{2}$, $c_s^2$ remains positive throughout the displayed temperature interval, consistent with a smooth crossover and the absence of spinodal branches. As the critical field is approached from the subcritical side,
the minimum of $c_s^2$ tends to zero, indicating the corresponding
softening of the equation of state. Sound speed is used here as a
thermodynamic diagnostic. The pseudocritical temperature is defined
from the entropy inflection in section~\ref{sec:phase_diagram}.
At the critical field, the minimum of $c_s^2$ reaches zero at the critical temperature $T_c$, where the entropy turning points coincide and $C_V$ diverges. The three criteria give the same CEP. Above the critical field, $c_s^2$ vanishes at the two spinodal temperatures, which delimit the multibranch interval. It is negative on
the middle branch and positive on the two outer branches. The identity $C_V=s/c_s^2$ relates these results. At the CEP, $c_s^2=0$
corresponds to divergent $C_V$. On the intermediate branch, $c_s^2<0$ gives
$C_V<0$ and hence local thermodynamic instability.

\subsection{Phase diagram, transition boundary, and the critical end point}
\label{sec:phase_diagram}

We apply the transition criteria to fixed-field slices over
$eB\in[0,10]\,\mathrm{GeV}^{2}$. In the crossover region, we define
$\mathcal S(T,eB)=s(T,eB)/T^3$ and locate the transition-region
inflection with the largest positive slope:
\begin{equation}
\left(\frac{\partial^2\mathcal S}{\partial T^2}\right)_{eB}=0,
\qquad
\left(\frac{\partial\mathcal S}{\partial T}\right)_{eB}>0.
\label{eq:entropy_inflection_Tpc}
\end{equation}
At the selected inflection, the positive first derivative has a local
maximum; if several candidates occur, we choose the largest positive
slope in the transition region. This defines $T_{\rm pc}(eB)$.
The minimum of $c_s^2$ remains a separate diagnostic of softening.
In the first-order region, the equal-pressure condition in
eq.~\eqref{eq:maxwell_section4} determines the transition
temperature \(T_t(eB)\). The pseudocritical line and the first-order transition line meet at the
critical end point, where the entropy turning points coincide and $C_V$
diverges.

\begin{figure}[tbp]
\centering
\begin{tikzpicture}
  \node[anchor=south west,inner sep=0] (phaseplot) at (0,0)
    {\includegraphics[width=0.98\linewidth]{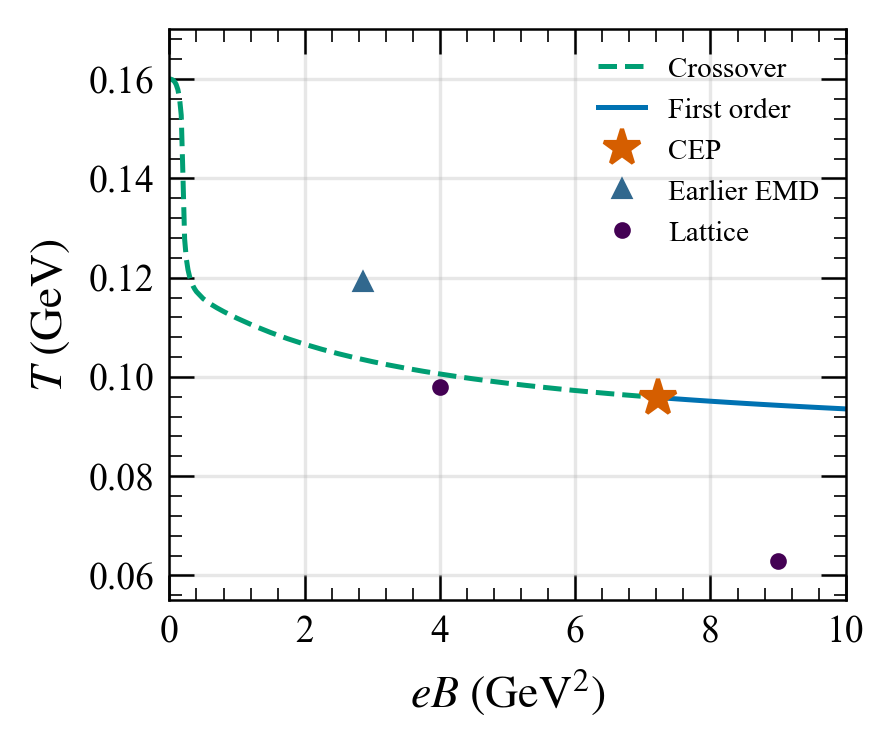}};
\end{tikzpicture}
\caption{Magnetic-field--temperature phase diagram of the recalibrated
model. The transition-region inflection of $s/T^3$ with the largest
positive slope defines the pseudocritical temperature $T_{\rm pc}(eB)$
in the crossover regime, while the equal-pressure condition defines
the first-order transition temperature $T_t(eB)$. These two definitions join continuously at the CEP, marked by the orange star at $(eB_c,T_c)\simeq(7.2200\,\mathrm{GeV}^{2},0.0958548\,\mathrm{GeV})$. The
blue triangle is the earlier EMD result~\cite{Cao:2024jgt}. Purple points are
the lattice crossover and first-order results at $eB=4$ and
$9\,\mathrm{GeV}^{2}$, respectively~\cite{DElia:2021yvk}.}
\label{fig:phase_diagram}
\end{figure}

Figure~\ref{fig:phase_diagram} shows the boundary obtained from these
thermodynamic criteria on the calibrated background. At zero field,
the entropy inflection gives
$T_{\rm pc}(0)\simeq0.16002\,\mathrm{GeV}$.
This temperature depends on the entropy-only calibration and the
chosen inflection criterion; it is not a separately fitted transition
temperature. The predicted boundary can be compared with the lattice
trend~\cite{Bali:2011qj,Bali:2014kia}. The entropy-only objective does
not directly fit transition temperatures, and the strong-field
extension includes temperatures below the $114\,\mathrm{MeV}$
training boundary.
The first-order transition temperature reaches
$T_t\simeq0.0935608\,\mathrm{GeV}$ at
$eB=10\,\mathrm{GeV}^{2}$.

Our CEP lies at a larger field than an earlier EMD result, which found
$eB_c\simeq2.8623\,\mathrm{GeV}^{2}$ and
$T_c\simeq0.1191\,\mathrm{GeV}$~\cite{Cao:2024jgt}.
Lattice calculations find a crossover at
$eB=4\,\mathrm{GeV}^{2}$ and a first-order transition at
$eB=9\,\mathrm{GeV}^{2}$~\cite{DElia:2021yvk}. The model reproduces this
qualitative change of transition order, with
$eB_c\simeq7.2200\,\mathrm{GeV}^{2}$ inside the interval
$4<eB_c<9\,\mathrm{GeV}^{2}$.

The transition temperature is less well reproduced: at
$eB=9\,\mathrm{GeV}^{2}$ the model gives
$T_t\simeq94.3\,\mathrm{MeV}$, compared with the lattice result
$63\pm5\,\mathrm{MeV}$~\cite{DElia:2021yvk}.

The first-order transition line extending to $eB=10\,\mathrm{GeV}^{2}$
is an extrapolation of the calibrated EMD model on the selected connected
solution family; the comparison with lattice results concerns the
qualitative change of order.

\section{Summary}
\label{sec:summary}

We have studied the magnetic-field-driven thermal transition at zero
baryon chemical potential by combining an anisotropic Einstein--Maxwell--dilaton
model with lattice calibration. The revised zero-field background
is calibrated to the zero-field entropy of ref.~\cite{Bali:2014kia}.
Sound speed follows from the same entropy curve, while pressure,
energy density, and trace anomaly use the stated pressure anchor.
Comparisons with the lattice equation of state provide conditional
consistency checks. The fit gives a deterministic estimate without
posterior uncertainties.

The finite-field calibration uses this same background. A positive,
monotonically decreasing $f(\phi)$ is determined from entropy data at
$eB=0.1$, $0.2$, $0.3$, and $0.4\,\mathrm{GeV}^2$, using the same
weighted-band strategy. Three neighbouring excluded fields are compared
before extrapolating to strong fields.

At low fields, the entropy density varies smoothly with temperature and
$C_V$ remains smooth and finite, consistent with a crossover. At the critical field,
the two entropy turning points coincide and $C_V$ diverges. Above this field, the turning points separate and produce an $S$-shaped structure in the thermodynamic functions, signalling the first-order regime. The minimum of $c_s^2$ reaches zero at
the same critical point. These signatures mutually confirm the CEP.

The magnetic field also changes the equation of state. With the common
fixed-field pressure subtraction, the energy density and trace anomaly
$I/T^4=(\varepsilon-3p)/T^4$ follow from the same entropy and pressure.
The minimum of $c_s^2$ tends to zero on approaching the critical
field from below.  Above the CEP,
the pressure develops a swallowtail.  The equal-pressure Maxwell construction
then fixes the transition temperature. The energy density and trace anomaly inherit the corresponding $S$-shaped structure of the thermodynamic quantities and provide a check of their thermodynamic consistency. The intermediate branch has $c_s^2<0$
and is locally thermodynamically unstable.

In the phase diagram, we define the pseudocritical temperature
\(T_{\rm pc}\) by the transition-region inflection of $s/T^3$ with
the largest positive slope. Sound-speed softening is a separate
thermodynamic diagnostic. The pseudocritical line joins continuously
with the equal-pressure first-order line at the CEP, giving a
connected transition curve in the magnetic-field--temperature plane.
This boundary is conditional on the entropy-only calibration and the
stated transition criteria and the assumed monotonicity of $f(\phi)$.
Our calibrated model locates the CEP at
\((eB_c,T_c)\simeq(7.2200\,\mathrm{GeV}^{2},0.0958548\,\mathrm{GeV})\). The
field coordinate lies within the lattice window
\(4<eB_c<9\,\mathrm{GeV}^{2}\)~\cite{DElia:2021yvk}, in qualitative
agreement with the lattice change of transition order.

\appendix
\section{Numerical method and calibration objective}
\label{app:architecture}

The zero-field calculation uses the isotropic six-state reduction of
the radial equations. The finite-field calculation evolves the
anisotropic eight-state system. Both are integrated with RK4 in
double precision; the ultraviolet coefficients determine the physical
observables. Both calibrations use
adaptive Nelder--Mead. The five independent background parameters are
fitted first, and the four coupling parameters are then fitted while
that background remains fixed.

\subsection{Radial evolution}
\label{app:RK4}

For the anisotropic calculation, the bulk state at each
radial point is specified by $a(r)$, $c(r)$, $h(r)$,
$\phi(r)$, and their radial derivatives. We introduce the derivative variables
\begin{equation}
\begin{aligned}
\pi_a&=a',&\pi_c&=c',\\
\pi_h&=h',&\pi_\phi&=\phi'.
\end{aligned}
\end{equation}
The equations of motion in section~\ref{sec:action} can then be written as
eight coupled first-order equations for
$\mathbf X(r)\equiv(a,c,h,\phi)$ and
$\bm{\pi}(r)\equiv(\pi_a,\pi_c,\pi_h,\pi_\phi)$. Following eq.~(12) of
ref.~\cite{Akutagawa:2020yeo}, their schematic one-step form is
\begin{equation}
\label{eq:discrete-EOM}
\left\{
\begin{aligned}
\mathbf X(r+\Delta r) &= \mathbf X(r) + \Delta r\,\bm{\pi}(r)\,,\\[2pt]
\bm{\pi}(r+\Delta r) &= \bm{\pi}(r) + \Delta r\,\mathcal{G}\!\bigl(\mathbf X(r),\bm{\pi}(r);\,V,f,\mathcal{B}\bigr)\,,
\end{aligned}
\right.
\end{equation}
where $\mathcal{G}$ denotes the expressions for
$(a'',c'',h'',\phi'')$ obtained from the second-order equations. At zero
field, $\mathcal{G}$ depends on $V(\phi)$. At finite field, it also depends
on $f(\phi)$ and $\mathcal B$. Equation~\eqref{eq:discrete-EOM} displays an
Euler step only to illustrate the layer structure. In the calculation, every
update is replaced by a fourth-order Runge--Kutta step. Its local truncation
error is $\mathcal O(\Delta r^5)$. The same finite set of model
parameters enters every radial update.

Iterating the RK4 map from $r=r_{\rm ini}$ to $r=r_{\rm max}$ gives
\begin{widetext}
\begin{equation}
\label{eq:DNN-map}
\begin{pmatrix}\mathbf X(r_{\rm max})\\ \bm{\pi}(r_{\rm max})\end{pmatrix}
= \mathcal{F}\!\bigl(\mathbf X(r_{\rm ini}),\,\bm{\pi}(r_{\rm ini});\,\phi_0,\,\mathcal{B};\,V(\phi),f(\phi)\bigr)\,.
\end{equation}
\end{widetext}
This map is the counterpart of eq.~(13) of
ref.~\cite{Akutagawa:2020yeo}. Each radial RK4 step acts as one network layer,
with the fields and their derivatives as its state variables.
The EMD equations determine the layer update, with the parameters of
$V(\phi)$ and $f(\phi)$ shared across radial steps. The normalisations
$\kappa^2$ and $\Lambda$ enter the thermodynamic dictionary.
Table~\ref{tab:EOM-DNN}
summarises this correspondence.

\begin{table*}[tp]
\centering
\renewcommand{\arraystretch}{1.2}
\begin{tabular}{lll}
\hline\hline
 & Zero field & Finite field \\
\hline
Horizon inputs & $\phi_0$ & $(\phi_0,\mathcal B)$ \\
Radial states & $(A,A',h,h',\phi,\phi')$ & $(\mathbf X,\bm\pi)$ \\
Free parameters & $v_1,v_3,v_4,\kappa^2,\Lambda$ & $F_0,k,c_2,c_3$ \\
Update method & Adaptive Nelder--Mead & Adaptive Nelder--Mead \\
\hline\hline
\end{tabular}
\caption{The two numerical implementations. RK4 radial evolution
is shared by both. In the revised zero-field fit, $v_2$ is derived
from the fixed ultraviolet mass; it is not an independent weight.
The neural-layer interpretation of the anisotropic map follows
ref.~\cite{Akutagawa:2020yeo}.}
\label{tab:EOM-DNN}
\end{table*}

For the updated isotropic solver, define
$\mathbf Y=(A,A',h,h',\phi,\phi')$. Its second-order equations are
\begin{equation}
\label{eq:zero_isotropic_rhs}
\begin{aligned}
A''&=-\frac{\phi'^2}{6},&
h''&=-4A'h',\\
\phi''&=\frac{V_{,\phi}}{h}
-\left(4A'+\frac{h'}{h}\right)\phi'.
\end{aligned}
\end{equation}
The Hamiltonian constraint is monitored in the nonsingular form
\begin{equation}
\label{eq:zero_constraint}
24hA'^2-h\phi'^2+6A'h'+2V=0.
\end{equation}
The Hamiltonian constraint is monitored in the ultraviolet region
as a solution-admission check, rather than a fitting target.
The integration starts from the near-horizon expansion and ends
in the asymptotic fitting region. A nonuniform radial grid resolves
the near-horizon evolution.

\begin{figure*}[t]
\centering
\includegraphics[width=0.32\textwidth]{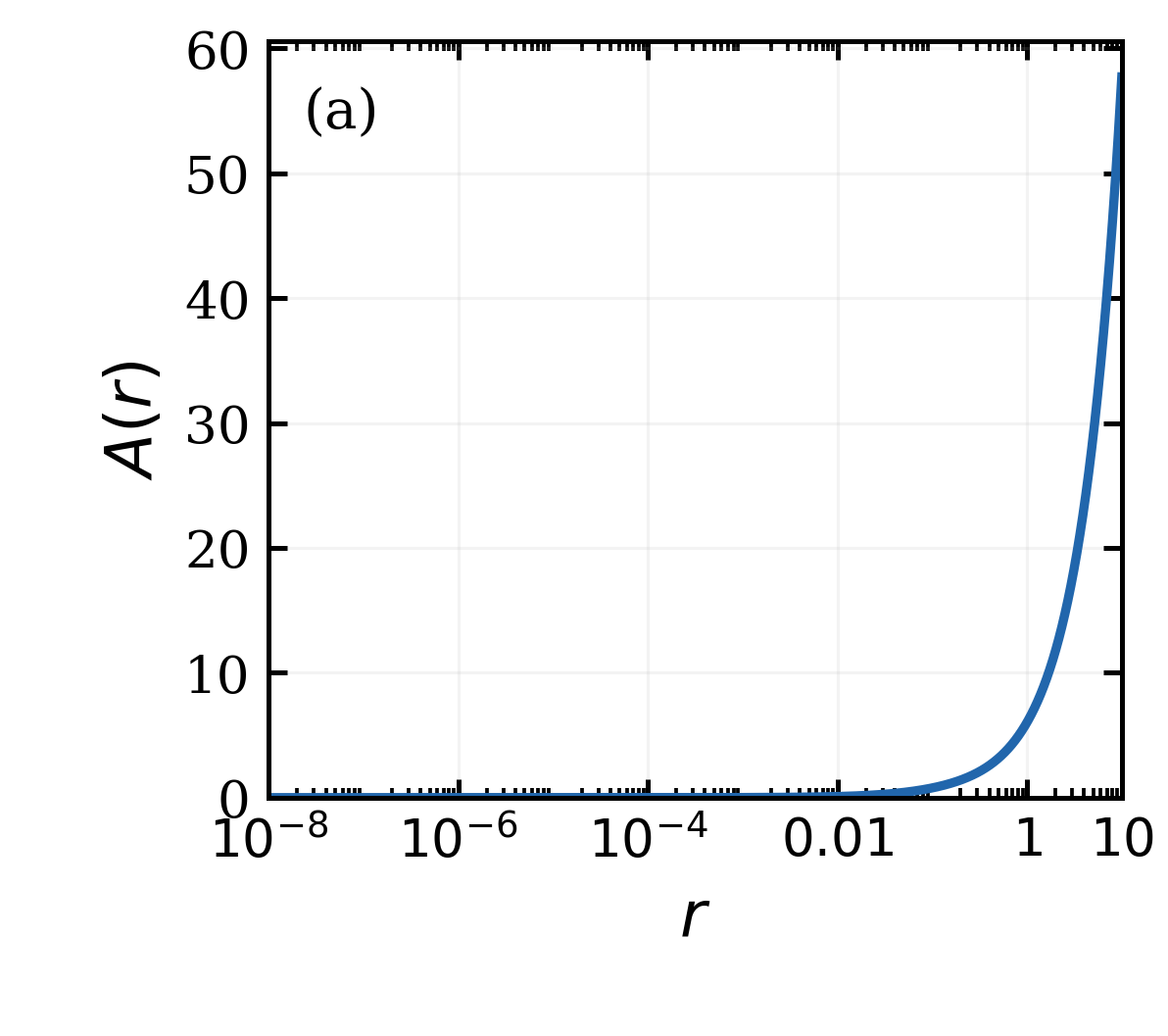}\hfill
\includegraphics[width=0.32\textwidth]{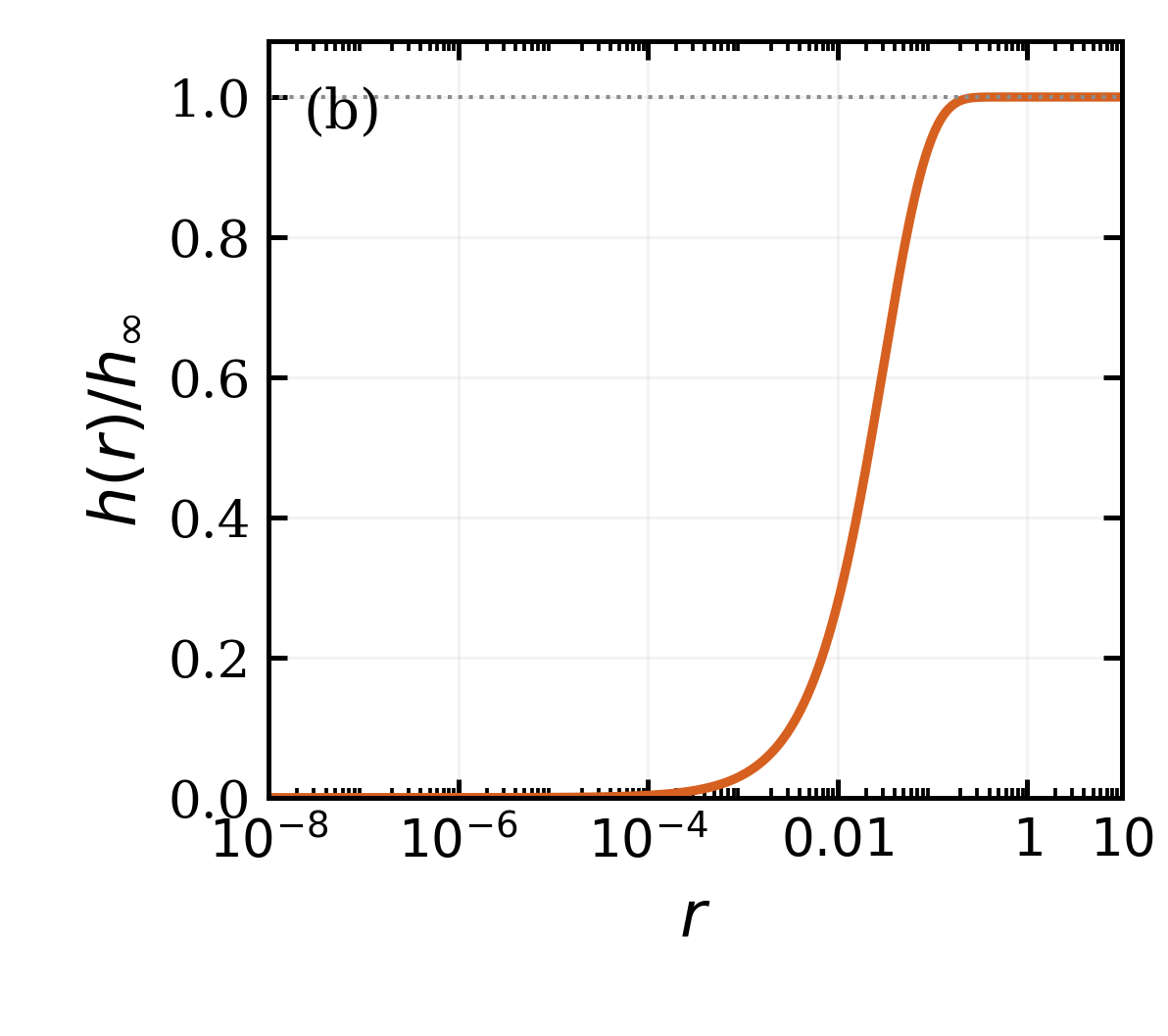}\hfill
\includegraphics[width=0.32\textwidth]{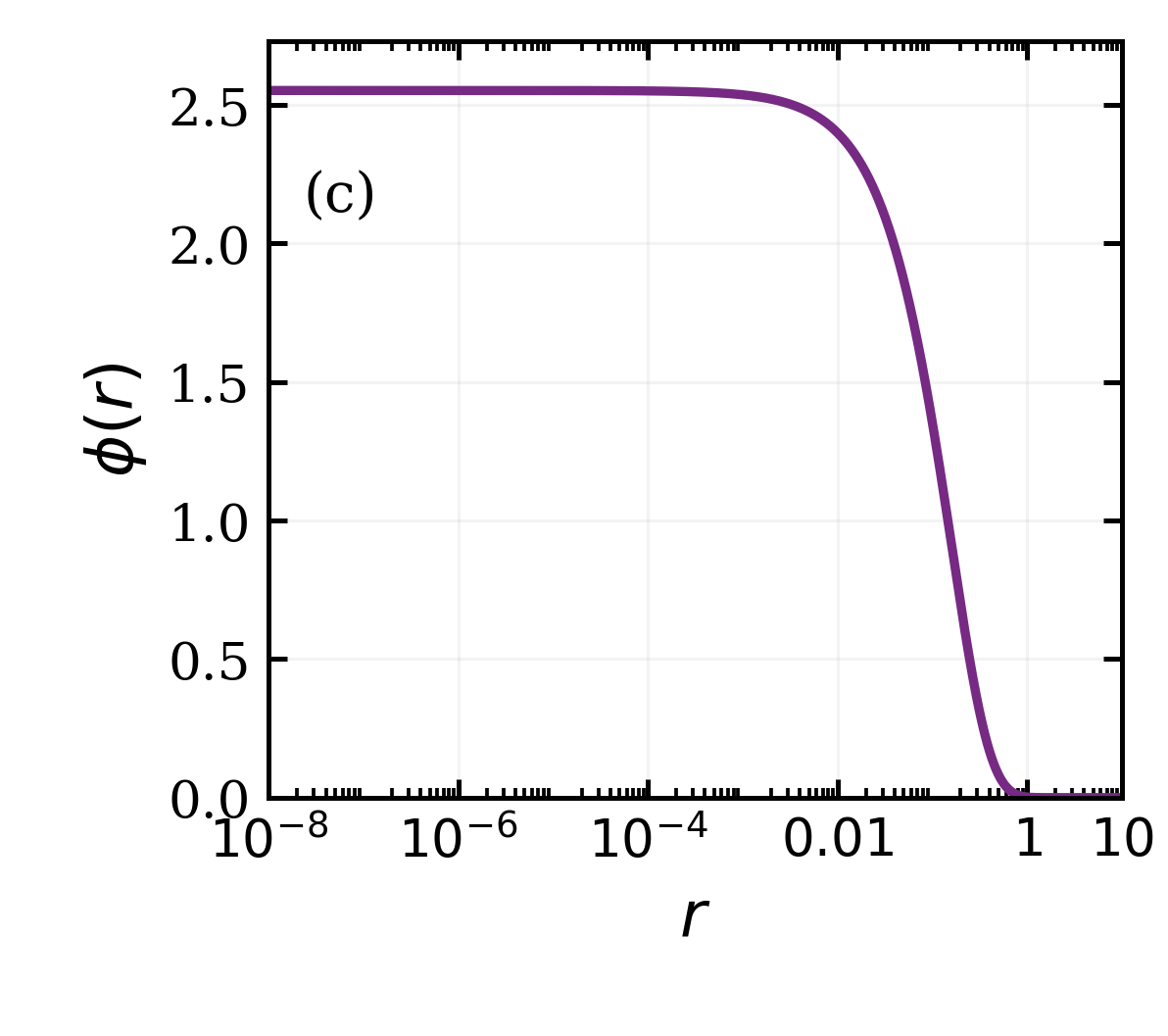}
\caption{Radial profiles of the recalibrated zero-field background at
$T=160\,\mathrm{MeV}$: (a) the common warp factor $A=a=c$,
(b) the blackening function normalised by
$h_\infty\equiv h_0^{\rm far}$, and (c) the dilaton.
The horizontal axes are logarithmic, with ticks labelled by $r$.
The horizon at $r_H=0$ lies outside the axes. The conformal boundary is at $r\to\infty$.
In the far region, the fields approach
$A'=1/\sqrt{h_\infty}$, $h=h_\infty$, and an exponentially decaying
dilaton.}
\label{fig:zero_radial_profiles}
\end{figure*}

For example, figure~\ref{fig:zero_radial_profiles} shows the recalibrated
zero-field solution at $T=160\,\mathrm{MeV}$.
At $eB=0$, $a=c=A$. The plotted fields are components of $\mathbf Y$
and follow from eq.~\eqref{eq:zero_isotropic_rhs} and the horizon data.
Thus, forward propagation integrates the bulk field equations.
The optimizer only updates the shared parameters; in the new
zero-field fit, these updates use Nelder--Mead.

\subsection{Input values at the initial layer}
\label{app:initial-layer}

The input layer is placed just outside the horizon at
$r=r_{\rm ini}$. Its eight components are the fields
$(a,c,h,\phi)$ and their derivative variables
$(\pi_a,\pi_c,\pi_h,\pi_\phi)$. Their initial values follow from a
near-horizon Taylor expansion of the EMD equations, generalising eq.~(14) of
ref.~\cite{Akutagawa:2020yeo}. For each field, we write
\begin{equation}
X(r) = X_0 + X_1 r + X_2 r^2 + \mathcal{O}(r^3)\,,\qquad
X\in\{a,c,h,\phi\}\,.
\end{equation}
Using the horizon gauge and coordinate normalisation of
eq.~\eqref{eq:horizon_gauge}, we set
\begin{equation}
a_0 = c_0 = h_0 = 0\,,\qquad h_1 = 1\,.
\end{equation}
Horizon regularity gives $h_0=0$. The conditions $a_0=c_0=0$ and
$h_1=h'(r_H)=1$ fix the residual coordinate normalisations. In these
coordinates, the Hawking temperature is $1/(4\pi)$. The physical temperature
is obtained after restoring the boundary-time normalisation and the scale
$\Lambda$.

At finite $\mathcal B$, the Maxwell sector produces anisotropic terms through
$f(\phi)$. The first-order coefficients in the horizon expansion are
\begin{equation}
\begin{aligned}
a_1 &= -\tfrac{1}{3}V_H + \tfrac{1}{6} f_{\rm H} \mathcal B^2\,,\\
c_1 &= -\tfrac{1}{3}V_H - \tfrac{1}{3} f_{\rm H} \mathcal B^2\,,\\
\phi_1 &= V'_H + \tfrac{1}{2} f'_{\rm H} \mathcal B^2\,,
\end{aligned}
\end{equation}
where $V_H\equiv V(\phi_0)$, $f_{\rm H}\equiv f(\phi_0)$, and
$f'_{\rm H}\equiv f'(\phi_0)$. Primes denote derivatives with respect to
$\phi$. The second-order coefficients have the form
\begin{equation}
\begin{aligned}
h_2 &= -(a_1 + c_1)\,,\\
a_2 &= a_2^{(0)} + a_2^{(2)}\mathcal B^2 + a_2^{(4)}\mathcal B^4\,,\\
c_2 &= c_2^{(0)} + c_2^{(2)}\mathcal B^2 + c_2^{(4)}\mathcal B^4\,,\\
\phi_2 &= \phi_2^{(0)} + \phi_2^{(2)}\mathcal B^2
          + \phi_2^{(4)}\mathcal B^4\,.
\end{aligned}
\end{equation}
The coefficients labelled by $(0)$, $(2)$, and $(4)$ depend on
$V(\phi_0)$, $V'(\phi_0)$, $V''(\phi_0)$, $f(\phi_0)$, $f'(\phi_0)$, and
$f''(\phi_0)$. They are determined algebraically from the EMD equations. In
the limit $\mathcal B\to0$, the anisotropic terms vanish and $a_1=c_1$. The
expansion then reduces to the isotropic Einstein--dilaton result.
For the updated six-state solver, its explicit coefficients are
\begin{equation}
\label{eq:zero_horizon_coefficients}
\begin{aligned}
 A_0&=h_0=0,& h_1&=1,\\
 A_1&=-V_H/3,& A_2&=-V_{,\phi,H}^{\,2}/12,\\
 h_2&=2V_H/3,& \phi_1&=V_{,\phi,H},\\
 \phi_2&=V_{,\phi,H}V_{,\phi\phi,H}/4.
\end{aligned}
\end{equation}
Here $X_2$ is the coefficient multiplying $r^2$, without a factorial.
Fields and derivatives at $r_{\rm ini}$ are evaluated from these
second-order expansions.

\subsection{Ultraviolet extraction and numerical checks}
\label{app:hyperparameters}

The scalar source and response are obtained from a column-scaled QR
least-squares fit to the two ultraviolet modes with fixed exponents
$\nu$ and $\Delta$. Metric intercepts and the asymptotic blackening
coefficient are extracted in the same far region. The thermodynamic
dictionary then gives the temperature, magnetic field, and entropy.
Regularity, positive leading scalar source, and positive temperature
and entropy are required for accepted solutions.

The model support adapts to cover the lattice temperature range.
A natural cubic entropy interpolant is evaluated at the reference
temperatures. Only the model is interpolated; the lattice centers
and reported bands remain fixed. Parameter sets that fail coverage
or the physical acceptance conditions are rejected.

Local convergence is assessed through simplex contraction, feasible
parameter perturbations, and stability under restarts. The selected
parameters are independently replayed with increased radial and
thermodynamic resolution. The ultraviolet fitting interval is varied
for a representative zero-field solution. These are local numerical checks;
they do not establish a global optimum, parameter uniqueness, or
uncertainties of the fitted model.

For strong-field predictions, the physical field is re-solved at
each $\phi_0$ on the regular family connected to $\mathcal B=0$.
The connected ordering is retained through thermal turning points;
disconnected solutions are excluded. Writing
$D=(dT/d\phi_0)_{eB}$, the chain rule gives
$\mathcal S_{,T}=\mathcal S_{,\phi_0}/D$ and
$\mathcal S_{,TT}=(\mathcal S_{,\phi_0\phi_0}D-
\mathcal S_{,\phi_0}D_{,\phi_0})/D^3$.
We select the transition-region inflection with the largest positive
$\mathcal S_{,T}$ for the crossover and solve $D=0$ and
$D_{,\phi_0}=0$ for the CEP. Root refinement and resolution changes
check these determinations. Pressure comparisons use a common
primitive along the connected curve.

\subsection{Loss function of the zero-field calibration}
\label{app:loss-V}

Only the $eB=0$ entropy group of ref.~\cite{Bali:2014kia} determines
the five independent background parameters. Define the fixed
tabulated centers $\mathcal S_i^{\rm lat}$, their half-widths
$\delta_i$, and
\begin{equation}
A_{\mathcal S}=\frac1N\sum_{i=1}^{N}|\mathcal S_i^{\rm lat}|,
\qquad
r_i=\mathcal S_\theta(T_i)-\mathcal S_i^{\rm lat}.
\end{equation}
Here $N$ is the number of reference temperatures.
The objective is
\begin{equation}
\label{eq:LtotV}
\begin{aligned}
\mathcal L_0&=\frac{1}{N A_{\mathcal S}}\sum_{i=1}^{N}
\ell(r_i,\delta_i),\\
\ell(r,\delta)&=\alpha |r|+(1-\alpha)\max(|r|-\delta,0),\\
\alpha&=0.05.
\end{aligned}
\end{equation}
The first term favors the center within the band, while the second
increases the penalty outside it. The slopes in
$|r_i|$ are $0.05$ inside and $1$ outside, up to the common
normalization. The objective is continuous and retains a nonzero slope
inside the band. It does not require every point to lie in its band
and is not a Gaussian likelihood normalized by the errors.

There is no sound-speed data term, pressure data term, derivative
regularizer, or additional shape penalty. We reject entropy splines that
violate stability or causality, rather than adding soft penalties. The
free-parameter signs and fixed ultraviolet mass relation are imposed
as described in section~\ref{sec:V_calib}.

The objective is minimised by bounded adaptive Nelder--Mead under
the stated physical constraints. Sound speed is calculated from
the same entropy interpolant and compared with lattice results
without entering the objective.

\subsection{Loss function of the coupling calibration}
\label{app:loss-f}

For each target field, the model entropy interpolant is evaluated
at the corresponding lattice temperatures. The fixed-field support
must cover the reference range. In strong-field predictions,
the $\phi_0$ ordering is retained through turning points.

Write $\mathcal S=s/T^3$, with residual
$r_{Bi}=\mathcal S_{\rm model}(T_i,B)-\mathcal S_{\rm lat}(T_i,B)$
and reported half-width $\delta_{Bi}$.
The finite-field objective is
\begin{equation}
\label{eq:LtotF}
\begin{aligned}
\mathcal L_B&=\frac{1}{|\mathcal D_4|}\sum_{B\in\mathcal D_4}
 \frac{1}{N_B A_B}\sum_{i=1}^{N_B}
 \ell(r_{Bi},\delta_{Bi}),\\
A_B&=\frac1{N_B}\sum_{i=1}^{N_B}|\mathcal S_{\rm lat}(T_i,B)|.
\end{aligned}
\end{equation}
where $\mathcal D_4=\{0.1,0.2,0.3,0.4\}\,\mathrm{GeV}^2$,
$N_B$ is the number of reference temperatures at field $B$, and
$\alpha=0.05$ as in the zero-field objective.
Each field has equal weight. Residuals inside the band retain a weak
central-value penalty, while residuals outside the band receive the
additional excess penalty. The construction follows the zero-field
weighted-band strategy in appendix~\ref{app:loss-V}.
There are no susceptibility, sound-speed, temperature-region, or
first- and second-difference terms. This deterministic objective does
not define a likelihood or yield a Bayesian posterior for correlated
temperature rows.

The complete coupling, rather than its separate terms, must obey
$f>0$ and $f'\leq0$. Put
\begin{equation}
\label{eq:f_polynomial_shape}
P(u)=(1+c_2-c_3)u+(-c_2+3c_3)u^2-2c_3u^3.
\end{equation}
For the $\phi\geq0$ scalar branch used here, $0<u\leq1$ and
$du/d\phi\leq0$. Positivity and monotonicity therefore follow from $P(0)=0$ and $P'(u)\geq0$ on $[0,1]$, with
$F_0,k>0$. For each parameter state, the extrema of the quadratic
$P'(u)$ are checked continuously, rather than only at sampled
$\phi$ values. The fitted coupling satisfies this condition for the complete
function.

\subsection{Initialisation}
\label{app:initial-weights}

The trainable weights require a stable initialisation, as also found
in ref.~\cite{Akutagawa:2020yeo}. For the zero-field fit, we start
from the previously fitted background candidate and perform local
refinement, as described in section~\ref{sec:V_calib}.
The coupling fit is refined from a feasible four-parameter pilot
solution of the same family, with $F_0$ and $k$ free and positive.
Every trial state must satisfy the continuous coupling-shape test
and the radial regularity checks. The fitted parameters are then
replayed independently at increased numerical resolution.

\subsection{Training dataset}
\label{app:dataset}

The background and coupling use the zero-field and finite-field
entropy, respectively, from ref.~\cite{Bali:2014kia}. The full
entropy is used rather than the magnetic increment. The reported
entropy and half-width are converted to $s/T^3$ at their respective
temperatures. The lattice observations are not augmented by
interpolation.

The coupling training fields are
$eB=0.1,0.2,0.3,0.4\,\mathrm{GeV}^2$; the fields
$eB=0.5,0.6,0.7\,\mathrm{GeV}^2$ are withheld for comparison.
WB denotes ref.~\cite{Borsanyi:2013bia}, and HotQCD denotes
ref.~\cite{HotQCD:2014kol}. These groups provide entropy and
sound-speed comparisons within the fitted temperature interval,
retaining asymmetric errors where applicable. None enters the final
background objective. These are conditional diagnostics, since
they share an equation-of-state source or were used in earlier fits.

\begin{acknowledgments}

This work is supported by the National Natural Science Foundation of China
(NSFC) under Grant Nos. 12305142 and 12275108, the Guangdong Basic and Applied
Basic Research Foundation under Grant No. 2024A1515012931, and the Science and
Technology Planning Project of Guangzhou, China under Grant No. 2024A04J3243.

\end{acknowledgments}

\bibliography{refs}

@article{Hippert:2023bel,
    author = "Hippert, Mauricio and Grefa, Joaquin and Manning, T. Andrew and Noronha, Jorge and Noronha-Hostler, Jacquelyn and Portillo Vazquez, Israel and Ratti, Claudia and Rougemont, Romulo and Trujillo, Michael",
    title = "{Bayesian location of the QCD critical point from a holographic perspective}",
    eprint = "2309.00579",
    archivePrefix = "arXiv",
    primaryClass = "nucl-th",
    doi = "10.1103/PhysRevD.110.094006",
    journal = "Phys. Rev. D",
    volume = "110",
    number = "9",
    pages = "094006",
    year = "2024"
}

@article{Rougemont:2023gfz,
    author = "Rougemont, Romulo and Grefa, Joaquin and Hippert, Mauricio and Noronha, Jorge and Noronha-Hostler, Jacquelyn and Portillo, Israel and Ratti, Claudia",
    title = "{Hot QCD phase diagram from holographic Einstein{\textendash}Maxwell{\textendash}Dilaton models}",
    eprint = "2307.03885",
    archivePrefix = "arXiv",
    primaryClass = "nucl-th",
    doi = "10.1016/j.ppnp.2023.104093",
    journal = "Prog. Part. Nucl. Phys.",
    volume = "135",
    pages = "104093",
    year = "2024"
}

@article{Wen:2024pion,
    author = "Wen, Nanxiang and Cao, Xuanmin and Chao, Jingyi and Liu, Hui",
    title = "{Neutral pion masses within a hot and magnetized medium in a lattice-improved soft-wall AdS/QCD model}",
    eprint = "2402.06239",
    archivePrefix = "arXiv",
    primaryClass = "hep-ph",
    doi = "10.1103/PhysRevD.109.086021",
    journal = "Phys. Rev. D",
    volume = "109",
    number = "8",
    pages = "086021",
    year = "2024"
}

@article{Fukushima:2010bq,
    author = "Fukushima, Kenji and Hatsuda, Tetsuo",
    title = "{The phase diagram of dense QCD}",
    eprint = "1005.4814",
    archivePrefix = "arXiv",
    primaryClass = "hep-ph",
    reportNumber = "YITP-10-28, TKYNT-10-06",
    doi = "10.1088/0034-4885/74/1/014001",
    journal = "Rept. Prog. Phys.",
    volume = "74",
    pages = "014001",
    year = "2011"
}

@article{Ratti:2018ksb,
    author = "Ratti, Claudia",
    title = "{Lattice QCD and heavy ion collisions: a review of recent progress}",
    eprint = "1804.07810",
    archivePrefix = "arXiv",
    primaryClass = "hep-lat",
    doi = "10.1088/1361-6633/aabb97",
    journal = "Rept. Prog. Phys.",
    volume = "81",
    number = "8",
    pages = "084301",
    year = "2018"
}

@article{Braun-Munzinger:2015hba,
    author = {Braun-Munzinger, Peter and Koch, Volker and Sch{\"a}fer, Thomas and Stachel, Johanna},
    title = "{Properties of hot and dense matter from relativistic heavy ion collisions}",
    eprint = "1510.00442",
    archivePrefix = "arXiv",
    primaryClass = "nucl-th",
    doi = "10.1016/j.physrep.2015.12.003",
    journal = "Phys. Rept.",
    volume = "621",
    pages = "76--126",
    year = "2016"
}

@article{Borsanyi:2010cj,
    author = "Borsanyi, Szabolcs and Endrodi, Gergely and Fodor, Zoltan and Jakovac, Antal and Katz, Sandor D. and Krieg, Stefan and Ratti, Claudia and Szabo, Kalman K.",
    title = "{The QCD equation of state with dynamical quarks}",
    eprint = "1007.2580",
    archivePrefix = "arXiv",
    primaryClass = "hep-lat",
    reportNumber = "WUB-10-12",
    doi = "10.1007/JHEP11(2010)077",
    journal = "JHEP",
    volume = "11",
    number = "11",
    pages = "077",
    year = "2010"
}

@article{HotQCD:2014kol,
    author = "Bazavov, A. and others",
    collaboration = "HotQCD",
    title = "{Equation of state in ( 2+1 )-flavor QCD}",
    eprint = "1407.6387",
    archivePrefix = "arXiv",
    primaryClass = "hep-lat",
    reportNumber = "BNL-105928-2014-JA",
    doi = "10.1103/PhysRevD.90.094503",
    journal = "Phys. Rev. D",
    volume = "90",
    pages = "094503",
    year = "2014"
}

@article{Borsanyi:2013bia,
    author = "Borsanyi, Szabolcs and Fodor, Zoltan and Hoelbling, Christian and Katz, Sandor D. and Krieg, Stefan and Szabo, Kalman K.",
    title = "{Full result for the QCD equation of state with 2+1 flavors}",
    eprint = "1309.5258",
    archivePrefix = "arXiv",
    primaryClass = "hep-lat",
    doi = "10.1016/j.physletb.2014.01.007",
    journal = "Phys. Lett. B",
    volume = "730",
    pages = "99--104",
    year = "2014"
}

@article{Bazavov:2011nk,
    author = "Bazavov, A. and others",
    title = "{The chiral and deconfinement aspects of the QCD transition}",
    eprint = "1111.1710",
    archivePrefix = "arXiv",
    primaryClass = "hep-lat",
    doi = "10.1103/PhysRevD.85.054503",
    journal = "Phys. Rev. D",
    volume = "85",
    pages = "054503",
    year = "2012"
}

@article{Skokov:2009qp,
    author = "Skokov, V. and Illarionov, A. Yu. and Toneev, V.",
    title = "{Estimate of the magnetic field strength in heavy-ion collisions}",
    eprint = "0907.1396",
    archivePrefix = "arXiv",
    primaryClass = "nucl-th",
    doi = "10.1142/S0217751X09047570",
    journal = "Int. J. Mod. Phys. A",
    volume = "24",
    pages = "5925--5932",
    year = "2009"
}

@article{Voronyuk:2011jd,
    author = "Voronyuk, V. and Toneev, V. D. and Cassing, W. and Bratkovskaya, E. L. and Konchakovski, V. P. and Voloshin, S. A.",
    title = "{(Electro-)Magnetic field evolution in relativistic heavy-ion collisions}",
    eprint = "1103.4239",
    archivePrefix = "arXiv",
    primaryClass = "nucl-th",
    doi = "10.1103/PhysRevC.83.054911",
    journal = "Phys. Rev. C",
    volume = "83",
    pages = "054911",
    year = "2011"
}

@article{Deng:2012pc,
    author = "Deng, Wei-Tian and Huang, Xu-Guang",
    title = "{Event-by-event generation of electromagnetic fields in heavy-ion collisions}",
    eprint = "1201.5108",
    archivePrefix = "arXiv",
    primaryClass = "nucl-th",
    doi = "10.1103/PhysRevC.85.044907",
    journal = "Phys. Rev. C",
    volume = "85",
    pages = "044907",
    year = "2012"
}

@article{Bzdak:2011yy,
    author = "Bzdak, Adam and Skokov, Vladimir",
    title = "{Event-by-event fluctuations of magnetic and electric fields in heavy ion collisions}",
    eprint = "1111.1949",
    archivePrefix = "arXiv",
    primaryClass = "hep-ph",
    reportNumber = "BNL-96541-2011-JA, RBRC-927",
    doi = "10.1016/j.physletb.2012.02.065",
    journal = "Phys. Lett. B",
    volume = "710",
    pages = "171--174",
    year = "2012"
}

@article{McLerran:2013hla,
    author = "McLerran, L. and Skokov, V.",
    title = "{Comments About the Electromagnetic Field in Heavy-Ion Collisions}",
    eprint = "1305.0774",
    archivePrefix = "arXiv",
    primaryClass = "hep-ph",
    reportNumber = "BNL-100762-2013-JA",
    doi = "10.1016/j.nuclphysa.2014.05.008",
    journal = "Nucl. Phys. A",
    volume = "929",
    pages = "184--190",
    year = "2014"
}

@article{Kharzeev:2012ph,
    author = "Kharzeev, Dmitri E. and Landsteiner, Karl and Schmitt, Andreas and Yee, Ho-Ung",
    title = "{'Strongly interacting matter in magnetic fields': an overview}",
    eprint = "1211.6245",
    archivePrefix = "arXiv",
    primaryClass = "hep-ph",
    doi = "10.1007/978-3-642-37305-3_1",
    journal = "Lect. Notes Phys.",
    volume = "871",
    pages = "1--11",
    year = "2013"
}

@article{Andersen:2014xxa,
    author = "Andersen, Jens O. and Naylor, William R. and Tranberg, Anders",
    title = "{Phase diagram of QCD in a magnetic field: A review}",
    eprint = "1411.7176",
    archivePrefix = "arXiv",
    primaryClass = "hep-ph",
    doi = "10.1103/RevModPhys.88.025001",
    journal = "Rev. Mod. Phys.",
    volume = "88",
    pages = "025001",
    year = "2016"
}

@article{Miransky:2015ava,
    author = "Miransky, Vladimir A. and Shovkovy, Igor A.",
    title = "{Quantum field theory in a magnetic field: From quantum chromodynamics to graphene and Dirac semimetals}",
    eprint = "1503.00732",
    archivePrefix = "arXiv",
    primaryClass = "hep-ph",
    doi = "10.1016/j.physrep.2015.02.003",
    journal = "Phys. Rept.",
    volume = "576",
    pages = "1--209",
    year = "2015"
}

@article{Huang:2015oca,
    author = "Huang, Xu-Guang",
    title = "{Electromagnetic fields and anomalous transports in heavy-ion collisions --- A pedagogical review}",
    eprint = "1509.04073",
    archivePrefix = "arXiv",
    primaryClass = "nucl-th",
    doi = "10.1088/0034-4885/79/7/076302",
    journal = "Rept. Prog. Phys.",
    volume = "79",
    number = "7",
    pages = "076302",
    year = "2016"
}

@article{Bazavov:2017dus,
    author = "Bazavov, A. and others",
    title = "{The QCD Equation of State to $\mathcal{O}(\mu_B^6)$ from Lattice QCD}",
    eprint = "1701.04325",
    archivePrefix = "arXiv",
    primaryClass = "hep-lat",
    doi = "10.1103/PhysRevD.95.054504",
    journal = "Phys. Rev. D",
    volume = "95",
    number = "5",
    pages = "054504",
    year = "2017"
}

@article{Bali:2011qj,
    author = "Bali, G. S. and Bruckmann, F. and Endrodi, G. and Fodor, Z. and Katz, S. D. and Krieg, S. and Schafer, A. and Szabo, K. K.",
    title = "{The QCD phase diagram for external magnetic fields}",
    eprint = "1111.4956",
    archivePrefix = "arXiv",
    primaryClass = "hep-lat",
    doi = "10.1007/JHEP02(2012)044",
    journal = "JHEP",
    volume = "02",
    number = "02",
    pages = "044",
    year = "2012"
}

@article{Bali:2014kia,
    author = {Bali, G. S. and Bruckmann, F. and Endr{\"o}di, G. and Katz, S. D. and Sch{\"a}fer, A.},
    title = "{The QCD equation of state in background magnetic fields}",
    eprint = "1406.0269",
    archivePrefix = "arXiv",
    primaryClass = "hep-lat",
    doi = "10.1007/JHEP08(2014)177",
    journal = "JHEP",
    volume = "08",
    number = "08",
    pages = "177",
    year = "2014"
}

@article{Bali:2013esa,
    author = "Bali, G. S. and Bruckmann, F. and Endrodi, G. and Gruber, F. and Schaefer, A.",
    title = "{Magnetic field-induced gluonic (inverse) catalysis and pressure (an)isotropy in QCD}",
    eprint = "1303.1328",
    archivePrefix = "arXiv",
    primaryClass = "hep-lat",
    doi = "10.1007/JHEP04(2013)130",
    journal = "JHEP",
    volume = "04",
    number = "04",
    pages = "130",
    year = "2013"
}

@article{Aoki:2006we,
    author = "Aoki, Y. and Endrodi, G. and Fodor, Z. and Katz, S. D. and Szabo, K. K.",
    title = "{The Order of the quantum chromodynamics transition predicted by the standard model of particle physics}",
    eprint = "hep-lat/0611014",
    archivePrefix = "arXiv",
    doi = "10.1038/nature05120",
    journal = "Nature",
    volume = "443",
    pages = "675--678",
    year = "2006"
}

@article{Bhattacharya:2014ara,
    author = "Bhattacharya, Tanmoy and others",
    title = "{QCD Phase Transition with Chiral Quarks and Physical Quark Masses}",
    eprint = "1402.5175",
    archivePrefix = "arXiv",
    primaryClass = "hep-lat",
    reportNumber = "BNL-103837-2014-JA, CU-TP-1205, INT-PUB-14-003, LLNL-JRNL-650194",
    doi = "10.1103/PhysRevLett.113.082001",
    journal = "Phys. Rev. Lett.",
    volume = "113",
    number = "8",
    pages = "082001",
    year = "2014"
}

@article{Borsanyi:2010bp,
    author = "Borsanyi, Szabolcs and Fodor, Zoltan and Hoelbling, Christian and Katz, Sandor D and Krieg, Stefan and Ratti, Claudia and Szabo, Kalman K.",
    collaboration = "Wuppertal-Budapest",
    title = "{Is there still any $T_c$ mystery in lattice QCD? Results with physical masses in the continuum limit III}",
    eprint = "1005.3508",
    archivePrefix = "arXiv",
    primaryClass = "hep-lat",
    reportNumber = "WUB-10-11, MIT-CTP-4152",
    doi = "10.1007/JHEP09(2010)073",
    journal = "JHEP",
    volume = "09",
    number = "09",
    pages = "073",
    year = "2010"
}

@article{Aoki:2009sc,
    author = "Aoki, Y. and Borsanyi, Szabolcs and Durr, Stephan and Fodor, Zoltan and Katz, Sandor D. and Krieg, Stefan and Szabo, Kalman K.",
    title = "{The QCD transition temperature: results with physical masses in the continuum limit II.}",
    eprint = "0903.4155",
    archivePrefix = "arXiv",
    primaryClass = "hep-lat",
    reportNumber = "WUB-09-01, ITP-BUDAPEST-644, RBRC-782",
    doi = "10.1088/1126-6708/2009/06/088",
    journal = "JHEP",
    volume = "06",
    number = "06",
    pages = "088",
    year = "2009"
}

@article{HotQCD:2019xnw,
    author = "Ding, H. T. and others",
    collaboration = "HotQCD",
    title = "{Chiral Phase Transition Temperature in ( 2+1 )-Flavor QCD}",
    eprint = "1903.04801",
    archivePrefix = "arXiv",
    primaryClass = "hep-lat",
    doi = "10.1103/PhysRevLett.123.062002",
    journal = "Phys. Rev. Lett.",
    volume = "123",
    number = "6",
    pages = "062002",
    year = "2019"
}

@article{Endrodi:2015oba,
    author = "Endrodi, Gergely",
    title = "{Critical point in the QCD phase diagram for extremely strong background magnetic fields}",
    eprint = "1504.08280",
    archivePrefix = "arXiv",
    primaryClass = "hep-lat",
    doi = "10.1007/JHEP07(2015)173",
    journal = "JHEP",
    volume = "07",
    number = "07",
    pages = "173",
    year = "2015"
}

@article{Maldacena:1997re,
    author = "Maldacena, Juan Martin",
    title = "{The Large $N$ limit of superconformal field theories and supergravity}",
    eprint = "hep-th/9711200",
    archivePrefix = "arXiv",
    reportNumber = "HUTP-97-A097, HUTP-98-A097",
    doi = "10.4310/ATMP.1998.v2.n2.a1",
    journal = "Adv. Theor. Math. Phys.",
    volume = "2",
    pages = "231--252",
    year = "1998"
}

@article{Witten:1998qj,
    author = "Witten, Edward",
    title = "{Anti de Sitter space and holography}",
    eprint = "hep-th/9802150",
    archivePrefix = "arXiv",
    reportNumber = "IASSNS-HEP-98-15",
    doi = "10.4310/ATMP.1998.v2.n2.a2",
    journal = "Adv. Theor. Math. Phys.",
    volume = "2",
    pages = "253--291",
    year = "1998"
}

@article{Gubser:1998bc,
    author = "Gubser, S. S. and Klebanov, Igor R. and Polyakov, Alexander M.",
    title = "{Gauge theory correlators from noncritical string theory}",
    eprint = "hep-th/9802109",
    archivePrefix = "arXiv",
    reportNumber = "PUPT-1767",
    doi = "10.1016/S0370-2693(98)00377-3",
    journal = "Phys. Lett. B",
    volume = "428",
    pages = "105--114",
    year = "1998"
}

@article{Gursoy:2007cb,
    author = "Gursoy, U. and Kiritsis, E.",
    title = "{Exploring improved holographic theories for QCD: Part I}",
    eprint = "0707.1324",
    archivePrefix = "arXiv",
    primaryClass = "hep-th",
    reportNumber = "CPHT-RR027-0507",
    doi = "10.1088/1126-6708/2008/02/032",
    journal = "JHEP",
    volume = "02",
    number = "02",
    pages = "032",
    year = "2008"
}

@article{Gursoy:2009jd,
    author = "Gursoy, U. and Kiritsis, E. and Mazzanti, L. and Nitti, F.",
    title = "{Improved Holographic Yang-Mills at Finite Temperature: Comparison with Data}",
    eprint = "0903.2859",
    archivePrefix = "arXiv",
    primaryClass = "hep-th",
    doi = "10.1016/j.nuclphysb.2009.05.017",
    journal = "Nucl. Phys. B",
    volume = "820",
    pages = "148--177",
    year = "2009"
}

@article{Gubser:2008ny,
    author = "Gubser, Steven S. and Nellore, Abhinav",
    title = "{Mimicking the QCD equation of state with a dual black hole}",
    eprint = "0804.0434",
    archivePrefix = "arXiv",
    primaryClass = "hep-th",
    reportNumber = "PUPT-2265",
    doi = "10.1103/PhysRevD.78.086007",
    journal = "Phys. Rev. D",
    volume = "78",
    pages = "086007",
    year = "2008"
}

@article{Gubser:2008yx,
    author = "Gubser, Steven S. and Nellore, Abhinav and Pufu, Silviu S. and Rocha, Fabio D.",
    title = "{Thermodynamics and bulk viscosity of approximate black hole duals to finite temperature quantum chromodynamics}",
    eprint = "0804.1950",
    archivePrefix = "arXiv",
    primaryClass = "hep-th",
    reportNumber = "PUPT-2266",
    doi = "10.1103/PhysRevLett.101.131601",
    journal = "Phys. Rev. Lett.",
    volume = "101",
    pages = "131601",
    year = "2008"
}

@article{DeWolfe:2010he,
    author = "DeWolfe, Oliver and Gubser, Steven S. and Rosen, Christopher",
    title = "{A holographic critical point}",
    eprint = "1012.1864",
    archivePrefix = "arXiv",
    primaryClass = "hep-th",
    reportNumber = "COLO-HEP-559, PUPT-2360",
    doi = "10.1103/PhysRevD.83.086005",
    journal = "Phys. Rev. D",
    volume = "83",
    pages = "086005",
    year = "2011"
}

@article{Finazzo:2014cna,
    author = "Finazzo, Stefano I. and Rougemont, Romulo and Marrochio, Hugo and Noronha, Jorge",
    title = "{Hydrodynamic transport coefficients for the non-conformal quark-gluon plasma from holography}",
    eprint = "1412.2968",
    archivePrefix = "arXiv",
    primaryClass = "hep-ph",
    doi = "10.1007/JHEP02(2015)051",
    journal = "JHEP",
    volume = "02",
    number = "02",
    pages = "051",
    year = "2015"
}

@article{Critelli:2017oub,
    author = "Critelli, Renato and Noronha, Jorge and Noronha-Hostler, Jacquelyn and Portillo, Israel and Ratti, Claudia and Rougemont, Romulo",
    title = "{Critical point in the phase diagram of primordial quark-gluon matter from black hole physics}",
    eprint = "1706.00455",
    archivePrefix = "arXiv",
    primaryClass = "nucl-th",
    doi = "10.1103/PhysRevD.96.096026",
    journal = "Phys. Rev. D",
    volume = "96",
    number = "9",
    pages = "096026",
    year = "2017"
}

@article{Knaute:2017opk,
    author = {Knaute, J. and Yaresko, R. and K{\"a}mpfer, B.},
    title = "{Holographic QCD phase diagram with critical point from Einstein{\textendash}Maxwell-dilaton dynamics}",
    eprint = "1702.06731",
    archivePrefix = "arXiv",
    primaryClass = "hep-ph",
    doi = "10.1016/j.physletb.2018.01.053",
    journal = "Phys. Lett. B",
    volume = "778",
    pages = "419--425",
    year = "2018"
}

@article{Grefa:2021qvt,
    author = "Grefa, Joaquin and Noronha, Jorge and Noronha-Hostler, Jacquelyn and Portillo, Israel and Ratti, Claudia and Rougemont, Romulo",
    title = "{Hot and dense quark-gluon plasma thermodynamics from holographic black holes}",
    eprint = "2102.12042",
    archivePrefix = "arXiv",
    primaryClass = "nucl-th",
    doi = "10.1103/PhysRevD.104.034002",
    journal = "Phys. Rev. D",
    volume = "104",
    number = "3",
    pages = "034002",
    year = "2021"
}

@article{Cai:2022omk,
    author = "Cai, Rong-Gen and He, Song and Li, Li and Wang, Yuan-Xu",
    title = "{Probing QCD critical point and induced gravitational wave by black hole physics}",
    eprint = "2201.02004",
    archivePrefix = "arXiv",
    primaryClass = "hep-th",
    doi = "10.1103/PhysRevD.106.L121902",
    journal = "Phys. Rev. D",
    volume = "106",
    number = "12",
    pages = "L121902",
    year = "2022"
}

@article{Rougemont:2015oea,
    author = "Rougemont, Romulo and Critelli, Renato and Noronha, Jorge",
    title = "{Holographic calculation of the QCD crossover temperature in a magnetic field}",
    eprint = "1505.07894",
    archivePrefix = "arXiv",
    primaryClass = "hep-th",
    doi = "10.1103/PhysRevD.93.045013",
    journal = "Phys. Rev. D",
    volume = "93",
    number = "4",
    pages = "045013",
    year = "2016"
}

@article{Gursoy:2016ofp,
    author = {G{\"u}rsoy, Umut and Iatrakis, Ioannis and J{\"a}rvinen, Matti and Nijs, Govert},
    title = "{Inverse Magnetic Catalysis from improved Holographic QCD in the Veneziano limit}",
    eprint = "1611.06339",
    archivePrefix = "arXiv",
    primaryClass = "hep-th",
    doi = "10.1007/JHEP03(2017)053",
    journal = "JHEP",
    volume = "03",
    number = "03",
    pages = "053",
    year = "2017"
}

@article{Bohra:2020qom,
    author = "Bohra, Hardik and Dudal, David and Hajilou, Ali and Mahapatra, Subhash",
    title = "{Chiral transition in the probe approximation from an Einstein-Maxwell-dilaton gravity model}",
    eprint = "2010.04578",
    archivePrefix = "arXiv",
    primaryClass = "hep-th",
    doi = "10.1103/PhysRevD.103.086021",
    journal = "Phys. Rev. D",
    volume = "103",
    number = "8",
    pages = "086021",
    year = "2021"
}

@article{Cao:2024jgt,
    author = "Cao, Xuanmin and Liu, Hui",
    title = "{Impact of the phase transition on quark-gluon plasma with an extremely strong magnetic field in holographic QCD}",
    eprint = "2408.00467",
    archivePrefix = "arXiv",
    primaryClass = "hep-th",
    doi = "10.1103/PhysRevD.111.026008",
    journal = "Phys. Rev. D",
    volume = "111",
    number = "2",
    pages = "026008",
    year = "2025"
}

@article{Cao:2020isospin,
    author = "Cao, Xuanmin and Liu, Hui and Li, Danning and Ou, Guanning",
    title = "{QCD phase diagram at finite isospin chemical potential and temperature in an IR-improved soft-wall AdS/QCD model}",
    eprint = "2001.02888",
    archivePrefix = "arXiv",
    primaryClass = "hep-ph",
    doi = "10.1088/1674-1137/44/8/083106",
    journal = "Chin. Phys. C",
    volume = "44",
    number = "8",
    pages = "083106",
    year = "2020"
}

@article{Cao:2020pion,
    author = "Cao, Xuanmin and Liu, Hui and Li, Danning",
    title = "{Pion quasiparticles and QCD phase transitions at finite temperature and isospin density from holography}",
    eprint = "2009.00289",
    archivePrefix = "arXiv",
    primaryClass = "hep-ph",
    doi = "10.1103/PhysRevD.102.126014",
    journal = "Phys. Rev. D",
    volume = "102",
    number = "12",
    pages = "126014",
    year = "2020"
}

@article{Hashimoto:2018ftp,
    author = "Hashimoto, Koji and Sugishita, Sotaro and Tanaka, Akinori and Tomiya, Akio",
    title = "{Deep learning and the AdS/CFT correspondence}",
    eprint = "1802.08313",
    archivePrefix = "arXiv",
    primaryClass = "hep-th",
    reportNumber = "OU-HET-961, RIKEN-iTHEMS-Report-18",
    doi = "10.1103/PhysRevD.98.046019",
    journal = "Phys. Rev. D",
    volume = "98",
    number = "4",
    pages = "046019",
    year = "2018"
}

@article{LeCun:2015pmr,
    author = "LeCun, Yann and Bengio, Yoshua and Hinton, Geoffrey",
    title = "{Deep learning}",
    doi = "10.1038/nature14539",
    journal = "Nature",
    volume = "521",
    pages = "436--444",
    year = "2015"
}

@article{Carleo:2019ptp,
    author = "Carleo, Giuseppe and Cirac, Ignacio and Cranmer, Kyle and Daudet, Laurent and Schuld, Maria and Tishby, Naftali and Vogt-Maranto, Leslie and Zdeborov{\'a}, Lenka",
    title = "{Machine learning and the physical sciences}",
    eprint = "1903.10563",
    archivePrefix = "arXiv",
    primaryClass = "physics.comp-ph",
    doi = "10.1103/RevModPhys.91.045002",
    journal = "Rev. Mod. Phys.",
    volume = "91",
    number = "4",
    pages = "045002",
    year = "2019"
}

@article{Hashimoto:2018bnb,
    author = "Hashimoto, Koji and Sugishita, Sotaro and Tanaka, Akinori and Tomiya, Akio",
    title = "{Deep Learning and Holographic QCD}",
    eprint = "1809.10536",
    archivePrefix = "arXiv",
    primaryClass = "hep-th",
    reportNumber = "RIKEN-iTHEMS-Report-18, OU-HET 996",
    doi = "10.1103/PhysRevD.98.106014",
    journal = "Phys. Rev. D",
    volume = "98",
    number = "10",
    pages = "106014",
    year = "2018"
}

@article{Akutagawa:2020yeo,
    author = "Akutagawa, Tetsuya and Hashimoto, Koji and Sumimoto, Takayuki",
    title = "{Deep Learning and AdS/QCD}",
    eprint = "2005.02636",
    archivePrefix = "arXiv",
    primaryClass = "hep-th",
    reportNumber = "OU-HET 1058",
    doi = "10.1103/PhysRevD.102.026020",
    journal = "Phys. Rev. D",
    volume = "102",
    number = "2",
    pages = "026020",
    year = "2020"
}

@article{Hashimoto:2019bih,
    author = "Hashimoto, Koji",
    title = "{AdS/CFT correspondence as a deep Boltzmann machine}",
    eprint = "1903.04951",
    archivePrefix = "arXiv",
    primaryClass = "hep-th",
    reportNumber = "OU-HET-1003",
    doi = "10.1103/PhysRevD.99.106017",
    journal = "Phys. Rev. D",
    volume = "99",
    number = "10",
    pages = "106017",
    year = "2019"
}

@article{Yan:2020wcd,
    author = "Yan, Yu-Kun and Wu, Shao-Feng and Ge, Xian-Hui and Tian, Yu",
    title = "{Deep learning black hole metrics from shear viscosity}",
    eprint = "2004.12112",
    archivePrefix = "arXiv",
    primaryClass = "hep-th",
    doi = "10.1103/PhysRevD.102.101902",
    journal = "Phys. Rev. D",
    volume = "102",
    number = "10",
    pages = "101902",
    month = "4",
    year = "2020"
}

@article{Chen:2024ckb,
    author = "Chen, Xun and Huang, Mei",
    title = "{Machine learning holographic black hole from lattice QCD equation of state}",
    eprint = "2401.06417",
    archivePrefix = "arXiv",
    primaryClass = "hep-ph",
    doi = "10.1103/PhysRevD.109.L051902",
    journal = "Phys. Rev. D",
    volume = "109",
    number = "5",
    pages = "L051902",
    year = "2024"
}

@article{Chen:2025goz,
    author = "Chen, Xun and Giannuzzi, Floriana and Nicotri, Stefano",
    title = "{Data-driven refinement of an analytical holographic model for the QCD phase transition}",
    eprint = "2512.01836",
    archivePrefix = "arXiv",
    primaryClass = "hep-ph",
    reportNumber = "BARI-TH/780-25",
    doi = "10.1103/j7wh-vf6x",
    journal = "Phys. Rev. D",
    volume = "113",
    number = "10",
    pages = "106007",
    year = "2026"
}

@article{Zhu:2025edv,
    author = "Zhu, Liqiang and Gao, Zhan and Ke, Weiyao and Zhang, Hanzhong",
    title = "{Bayesian inference of the magnetic field and chemical potential on holographic jet quenching in heavy-ion collisions}",
    eprint = "2506.00340",
    archivePrefix = "arXiv",
    primaryClass = "hep-ph",
    doi = "10.1088/1674-1137/ae2f4f",
    journal = "Chin. Phys. C",
    volume = "50",
    number = "4",
    pages = "044108",
    year = "2026"
}

@article{DHoker:2009mmn,
    author = "D'Hoker, Eric and Kraus, Per",
    title = "{Magnetic Brane Solutions in AdS}",
    eprint = "0908.3875",
    archivePrefix = "arXiv",
    primaryClass = "hep-th",
    doi = "10.1088/1126-6708/2009/10/088",
    journal = "JHEP",
    volume = "10",
    number = "10",
    pages = "088",
    year = "2009"
}

@article{Breitenlohner:1982bm,
    author = "Breitenlohner, Peter and Freedman, Daniel Z.",
    title = "{Stability in Gauged Extended Supergravity}",
    doi = "10.1016/0003-4916(82)90116-6",
    journal = "Annals Phys.",
    volume = "144",
    pages = "249",
    year = "1982"
}

@article{Breitenlohner:1982jf,
    author = "Breitenlohner, Peter and Freedman, Daniel Z.",
    title = "{Positive Energy in anti-De Sitter Backgrounds and Gauged Extended Supergravity}",
    reportNumber = "PRINT-82-0420 (MIT)",
    doi = "10.1016/0370-2693(82)90643-8",
    journal = "Phys. Lett. B",
    volume = "115",
    pages = "197--201",
    year = "1982"
}

@article{Klebanov:1999tb,
    author = "Klebanov, Igor R. and Witten, Edward",
    title = "{AdS / CFT correspondence and symmetry breaking}",
    eprint = "hep-th/9905104",
    archivePrefix = "arXiv",
    reportNumber = "PUPT-1863, IASSNS-HEP-99-49",
    doi = "10.1016/S0550-3213(99)00387-9",
    journal = "Nucl. Phys. B",
    volume = "556",
    pages = "89--114",
    year = "1999"
}

@misc{Zhu:2026dmc,
    author = "Zhu, De-Xing and Zhu, Li-Qiang and Chen, Xun and Hou, De-Fu and Zhou, Kai",
    title = "{Neural-Network Holographic Model of the QCD Phase Transition under Lattice and HRG Constraints}",
    eprint = "2603.25571",
    archivePrefix = "arXiv",
    primaryClass = "hep-ph",
    month = "3",
    year = "2026"
}

@article{Bekenstein:1973ur,
    author = "Bekenstein, Jacob D.",
    title = "{Black holes and entropy}",
    doi = "10.1103/PhysRevD.7.2333",
    journal = "Phys. Rev. D",
    volume = "7",
    pages = "2333--2346",
    year = "1973"
}

@article{Hawking:1975vcx,
    author = "Hawking, S. W.",
    editor = "Gibbons, G. W. and Hawking, S. W.",
    title = "{Particle Creation by Black Holes}",
    doi = "10.1007/BF02345020",
    journal = "Commun. Math. Phys.",
    volume = "43",
    pages = "199--220",
    year = "1975",
    note = "[Erratum: Commun.Math.Phys. 46, 206 (1976)]"
}

@article{Chamblin:1999tk,
    author = "Chamblin, Andrew and Emparan, Roberto and Johnson, Clifford V. and Myers, Robert C.",
    title = "{Charged AdS black holes and catastrophic holography}",
    eprint = "hep-th/9902170",
    archivePrefix = "arXiv",
    reportNumber = "DAMTP-1999-29, EHU-FT-9902, UK-99-02, MCGILL-99-07",
    doi = "10.1103/PhysRevD.60.064018",
    journal = "Phys. Rev. D",
    volume = "60",
    pages = "064018",
    year = "1999"
}

@article{Kubiznak:2012wp,
    author = "Kubiznak, David and Mann, Robert B.",
    title = "{P-V criticality of charged AdS black holes}",
    eprint = "1205.0559",
    archivePrefix = "arXiv",
    primaryClass = "hep-th",
    doi = "10.1007/JHEP07(2012)033",
    journal = "JHEP",
    volume = "07",
    number = "07",
    pages = "033",
    year = "2012"
}

@article{Vachaspati:1991nm,
    author = "Vachaspati, T.",
    title = "{Magnetic fields from cosmological phase transitions}",
    doi = "10.1016/0370-2693(91)90051-Q",
    journal = "Phys. Lett. B",
    volume = "265",
    pages = "258--261",
    year = "1991"
}

@article{DElia:2021yvk,
    author = "D'Elia, Massimo and Maio, Lorenzo and Sanfilippo, Francesco and Stanzione, Alfredo",
    title = "{Phase diagram of QCD in a magnetic background}",
    eprint = "2111.11237",
    archivePrefix = "arXiv",
    primaryClass = "hep-lat",
    doi = "10.1103/PhysRevD.105.034511",
    journal = "Phys. Rev. D",
    volume = "105",
    number = "3",
    pages = "034511",
    year = "2022"
}

@article{Li:2016gfn,
    author = "Li, Danning and Huang, Mei and Yang, Yi and Yuan, Pei-Hung",
    title = "{Inverse Magnetic Catalysis in the Soft-Wall Model of AdS/QCD}",
    eprint = "1610.04618",
    archivePrefix = "arXiv",
    primaryClass = "hep-th",
    doi = "10.1007/JHEP02(2017)030",
    journal = "JHEP",
    volume = "02",
    number = "02",
    pages = "030",
    year = "2017"
}

@article{Cai:2024eqa,
    author = "Cai, Rong-Gen and He, Song and Li, Li and Zeng, Hong-An",
    title = "{Neural ordinary differential equations for mapping the magnetic QCD phase diagram via holography}",
    eprint = "2406.12772",
    archivePrefix = "arXiv",
    primaryClass = "hep-th",
    doi = "10.1007/s11433-025-2914-x",
    journal = "Sci. China Phys. Mech. Astron.",
    volume = "69",
    number = "4",
    pages = "240414",
    year = "2026"
}

@article{Grasso:2000wj,
    author = "Grasso, Dario and Rubinstein, Hector R.",
    title = "{Magnetic fields in the early universe}",
    eprint = "astro-ph/0009061",
    archivePrefix = "arXiv",
    primaryClass = "astro-ph",
    doi = "10.1016/S0370-1573(00)00110-1",
    journal = "Phys. Rept.",
    volume = "348",
    pages = "163--266",
    year = "2001"
}

@article{Durrer:2013pga,
    author = "Durrer, Ruth and Neronov, Andrii",
    title = "{Cosmological Magnetic Fields: Their Generation, Evolution and Observation}",
    eprint = "1303.7121",
    archivePrefix = "arXiv",
    primaryClass = "astro-ph.CO",
    doi = "10.1007/s00159-013-0062-7",
    journal = "Astron. Astrophys. Rev.",
    volume = "21",
    pages = "62",
    year = "2013"
}

@article{Fang:2016qss,
    author = "Fang, Zhen",
    title = "{Anomalous dimension, chiral phase transition and inverse magnetic catalysis in soft-wall AdS/QCD}",
    doi = "10.1016/j.physletb.2016.04.044",
    journal = "Phys. Lett. B",
    volume = "758",
    pages = "1--8",
    year = "2016"
}

@article{Chen:2021mpt,
    author = "Chen, Xun and Zhang, Lin and Hou, Defu",
    title = "{Running coupling constant at finite chemical potential and magnetic field from holography}",
    eprint = "2108.03840",
    archivePrefix = "arXiv",
    primaryClass = "hep-ph",
    doi = "10.1088/1674-1137/ac5c2d",
    journal = "Chin. Phys. C",
    volume = "46",
    number = "7",
    pages = "073101",
    year = "2022"
}

@article{Cai:2012xh,
    author = "Cai, Rong-Gen and He, Song and Li, Danning",
    title = "{A hQCD model and its phase diagram in Einstein-Maxwell-Dilaton system}",
    eprint = "1201.0820",
    archivePrefix = "arXiv",
    primaryClass = "hep-th",
    doi = "10.1007/JHEP03(2012)033",
    journal = "JHEP",
    volume = "03",
    number = "03",
    pages = "033",
    year = "2012"
}

@article{Zhu:2023mfd,
    author = "Zhu, Zhou-Run and Hou, Defu",
    title = "{Inverse magnetic catalysis and energy loss in a holographic QCD model}",
    eprint = "2305.12375",
    archivePrefix = "arXiv",
    primaryClass = "hep-ph",
    doi = "10.1103/PhysRevD.110.066010",
    journal = "Phys. Rev. D",
    volume = "110",
    number = "6",
    pages = "066010",
    year = "2024"
}

@article{Li:2012ay,
    author = "Li, Danning and Huang, Mei and Yan, Qi-Shu",
    title = "{A dynamical soft-wall holographic QCD model for chiral symmetry breaking and linear confinement}",
    eprint = "1206.2824",
    archivePrefix = "arXiv",
    primaryClass = "hep-th",
    doi = "10.1140/epjc/s10052-013-2615-3",
    journal = "Eur. Phys. J. C",
    volume = "73",
    pages = "2615",
    year = "2013"
}

@article{Chen:2019rez,
    author = "Chen, Xun and Li, Danning and Hou, Defu and Huang, Mei",
    title = "{Quarkyonic phase from quenched dynamical holographic QCD model}",
    eprint = "1908.02000",
    archivePrefix = "arXiv",
    primaryClass = "hep-ph",
    doi = "10.1007/JHEP03(2020)073",
    journal = "JHEP",
    volume = "03",
    number = "03",
    pages = "073",
    year = "2020"
}

@article{Finazzo:2016mhm,
    author = "Finazzo, Stefano Ivo and Critelli, Renato and Rougemont, Romulo and Noronha, Jorge",
    title = "{Momentum transport in strongly coupled anisotropic plasmas in the presence of strong magnetic fields}",
    eprint = "1605.06061",
    archivePrefix = "arXiv",
    primaryClass = "hep-ph",
    doi = "10.1103/PhysRevD.94.054020",
    journal = "Phys. Rev. D",
    volume = "94",
    number = "5",
    pages = "054020",
    year = "2016",
    note = "[Erratum: Phys.Rev.D 96, 019903 (2017)]"
}

@article{LiHuang2013Spectra,
  author = {Li, Danning and Huang, Mei},
  title = {{Dynamical holographic QCD model for glueball and light meson spectra}},
  journal = {JHEP},
  volume = {11},
  pages = {088},
  year = {2013},
  doi = {10.1007/JHEP11(2013)088},
  eprint = {1303.6929},
  archivePrefix = {arXiv},
  primaryClass = {hep-ph},
  url = {https://inspirehep.net/literature/1225680}
}

@article{Chelabi2016SoftWall,
  author = {Chelabi, Kaddour and Fang, Zhen and Huang, Mei and Li, Danning and Wu, Yue-Liang},
  title = {{Chiral Phase Transition in the Soft-Wall Model of AdS/QCD}},
  journal = {JHEP},
  volume = {04},
  pages = {036},
  year = {2016},
  doi = {10.1007/JHEP04(2016)036},
  eprint = {1512.06493},
  archivePrefix = {arXiv},
  primaryClass = {hep-ph},
  url = {https://inspirehep.net/literature/1410851}
}

@article{LiEtAl2018CEP,
  author = {Li, Zhibin and Chen, Yidian and Li, Danning and Huang, Mei},
  title = {{Locating the QCD critical end point through the peaked baryon number susceptibilities along the freeze-out line}},
  journal = {Chin. Phys. C},
  volume = {42},
  pages = {013103},
  year = {2018},
  doi = {10.1088/1674-1137/42/1/013103},
  eprint = {1706.02238},
  archivePrefix = {arXiv},
  primaryClass = {hep-ph},
  url = {https://inspirehep.net/literature/1603096}
}

@article{ChenEtAl2019Criticality,
  author = {Chen, Xun and Li, Danning and Huang, Mei},
  title = {{Criticality of QCD in a holographic QCD model with critical end point}},
  journal = {Chin. Phys. C},
  volume = {43},
  pages = {023105},
  year = {2019},
  doi = {10.1088/1674-1137/43/2/023105},
  eprint = {1810.02136},
  archivePrefix = {arXiv},
  primaryClass = {hep-ph},
  url = {https://inspirehep.net/literature/1696981}
}

@article{ZhaoEtAl2024TwoFlavor,
  author = {Zhao, Yan-Qing and He, Song and Hou, Defu and Li, Li and Li, Zhibin},
  title = {{Phase structure and critical phenomena in two-flavor QCD by holography}},
  journal = {Phys. Rev. D},
  volume = {109},
  pages = {086015},
  year = {2024},
  doi = {10.1103/PhysRevD.109.086015},
  eprint = {2310.13432},
  archivePrefix = {arXiv},
  primaryClass = {hep-ph},
  url = {https://inspirehep.net/literature/2713110}
}

@article{JokelaEtAl2024Refining,
  author = {Jokela, Niko and J{\"a}rvinen, Matti and Piispa, Aleksi},
  title = {{Refining holographic models of the quark-gluon plasma}},
  journal = {Phys. Rev. D},
  volume = {110},
  pages = {126013},
  year = {2024},
  doi = {10.1103/PhysRevD.110.126013},
  eprint = {2405.02394},
  archivePrefix = {arXiv},
  primaryClass = {hep-th},
  url = {https://inspirehep.net/literature/2783544}
}

@article{EvansEtAl2016IMC,
  author = {Evans, Nick and Miller, Carlisson and Scott, Marc},
  title = {{Inverse Magnetic Catalysis in Bottom-Up Holographic QCD}},
  journal = {Phys. Rev. D},
  volume = {94},
  pages = {074034},
  year = {2016},
  doi = {10.1103/PhysRevD.94.074034},
  eprint = {1604.06307},
  archivePrefix = {arXiv},
  primaryClass = {hep-ph},
  url = {https://inspirehep.net/literature/1450002}
}

@article{GursoyEtAl2018BMu,
  author = {G{\"u}rsoy, Umut and J{\"a}rvinen, Matti and Nijs, Govert},
  title = {{Holographic QCD in the Veneziano Limit at a Finite Magnetic Field and Chemical Potential}},
  journal = {Phys. Rev. Lett.},
  volume = {120},
  pages = {242002},
  year = {2018},
  doi = {10.1103/PhysRevLett.120.242002},
  eprint = {1707.00872},
  archivePrefix = {arXiv},
  primaryClass = {hep-th},
  url = {https://inspirehep.net/literature/1608618}
}

@article{BohraEtAl2020Magnetic,
  author = {Bohra, Hardik and Dudal, David and Hajilou, Ali and Mahapatra, Subhash},
  title = {{Anisotropic string tensions and inversely magnetic catalyzed deconfinement from a dynamical AdS/QCD model}},
  journal = {Phys. Lett. B},
  volume = {801},
  pages = {135184},
  year = {2020},
  doi = {10.1016/j.physletb.2019.135184},
  eprint = {1907.01852},
  archivePrefix = {arXiv},
  primaryClass = {hep-th},
  url = {https://inspirehep.net/literature/1742576}
}

\end{document}